\documentclass[superscriptaddress,twocolumn,showpacs]{revtex4}
\usepackage{epsfig}

\newcommand{\simle}
{\raisebox{-0.75ex}[-1.5ex]{$\;\stackrel{<}{\sim}\;$}}

\def\d{{\partial}}
\def\s{{\sigma}}
\def\e{{\epsilon}}
\def\k{{ {\bf k} }}

\def\q{{ {\bf q} }}
\def\Q{{ {\bf Q} }}

\def\w{{\omega}}
\def\a{{\alpha}}
\def\b{{\beta}}

\begin{document}

\def\runtitle{
Effect of Nonmagnetic Impurity in Nearly Antiferromagnetic Fermi Liquid 
}
\def\runauthor
{
Hiroshi {\sc Kontani} and Masanori {\sc Ohno}
}

\title{
Effect of Nonmagnetic Impurity in Nearly Antiferromagnetic Fermi Liquid:\\
Magnetic Correlations and Transport Phenomena
}

\author{
Hiroshi {\sc Kontani} and Masanori {\sc Ohno}
}

\address{
Department of Physics, Nagoya University,
Furo-cho, Nagoya 464-8602, Japan.
}

\date{\today}

\begin{abstract}
In nearly antiferromagnetic (AF) metals
such as high-$T_{\rm c}$ superconductors (HTSC's),
a single nonmagnetic impurity frequently causes nontrivial widespread 
change of the electronic states.
To elucidate this long-standing issue, we study a Hubbard model 
with a strong onsite impurity potential based on an improved 
fluctuation-exchange (FLEX) approximation,
which we call the $GV^I$-FLEX method.
This model corresponds to the HTSC 
with dilute nonmagnetic impurity concentration.
We find that (i) both local and staggered susceptibilities
are strongly enhanced around the impurity.
By this reason, (ii) the quasiparticle lifetime as well as the
local density of states (DOS) are strongly suppressed 
in a wide area around the impurity (like a Swiss cheese hole), 
which causes the ``huge residual resistivity'' beyond the s-wave 
unitary scattering limit.
We stress that the excess quasiparticle damping rate
caused by impurities has strong $\k$-dependence due to 
non-s-wave scatterings induced by many-body effects,
so the structure of the ``hot spot/cold spot''
in the host system persists against impurity doping.
This result could be examined by the ARPES measurements.
In addition, (iii) only a few percent of impurities can causes
a ``Kondo-like'' upturn of resistivity ($d\rho/dT<0$) at low $T$ 
when the system is very close to the AF quantum critical point (QCP).
The results (i)-(iii) obtained in the present study,
which cannot be derived by the simple FLEX approximation,
naturally explains the main impurity effects in HTSC's.
We also discuss the impurity effect in
heavy fermion systems and organic superconductors.
\end{abstract}


\sloppy

\maketitle

\narrowtext

\section{Introduction}
\label{sec:intro}
In strongly correlated electron systems,
the presence of nonmagnetic impurities with low concentration
can cause drastic changes of electronic properties of the bulk system.
Thus, the impurity effect is a useful probe
to investigate the electronic states of the host system.
In under-doped high-$T_{\rm c}$ superconductors (HTSC's),
nonmagnetic impurities (such as Zn) causes a huge residual resistivity
beyond the s-wave unitary scattering limit.
Moreover, NMR measurements reveal that 
both the local and the staggered spin 
susceptibilities are strongly enhanced around the impurity.
Until now, the whole understanding of these impurity effect in HTSC
have not be achieved in terms of the Fermi liquid theory.
By this reason, nontrivial impurity effects in under-doped HTSC's 
are frequently considered as the evidence of the breakdown of the 
Fermi liquid state.
However, similar impurity effects are observed in other strongly
correlated metals such as heavy fermion (HF) systems or organic 
superconductors, near the magnetic quantum-critical-points (QCP).
Therefore, we have to develop previous theories of 
impurity effect to see whether these experimental results 
can be explained in terms of the Fermi liquid theory or not.

In HTSC's without impurities, 
various physical quantities in the normal state
deviate from the conventional Fermi liquid behaviors in usual metals,
which are called the non-Fermi liquid (NFL) behaviors.
Famous examples of the NFL behaviors would be 
the Curie-Weiss like behavior of $1/T_1T$ and the $T$-linear 
resistivity above the pseudo-gap temperature, $T^\ast \sim 200$K.
One of the most predominant candidates is the Fermi liquid state 
with strong antiferromagnetic (AF) fluctuations
 \cite{Yamada-rev,Moriya,Manske,Pines,Kontani-rev}.
In fact, spin fluctuation theories like the SCR theory
 \cite{Moriya} 
and the fluctuation-exchange (FLEX) approximation
 \cite{Bickers,Manske,Monthoux-Scalapino,Takimoto,Koikegami,Wermbter}
have succeeded in explaining various NFL behaviors in a unified way.
Recently, low-energy excitations in connection with HTSC's were studied 
in ref. 
 \cite{Manske-PRB}.
These approximations satisfy the Mermin-Wagner theorem (see Appendix A).
Moreover, pseudo-gap phenomena under $T^\ast$ are well reproduced
by the FLEX+$T$-matrix approximation, where self-energy correction 
due to strong superconducting (SC) fluctuations are taken into account
 \cite{Yamada-rev,Dahm-T,Kontani-rev}.

One of the most remarkable NFL behaviors in HTSC's 
would be the anomalous transport phenomena.
For example, the Hall coefficient $R_{\rm H}$ is proportional to 
$T^{-1}$ above $T^\ast$, and $|R_{\rm H}|\gg 1/ne$
($n$ being the electron filling number) at low $T$
  \cite{Satoh}.
Moreover, the magnetoresistance $\Delta\rho/\rho_0$ is 
proportional to $R_{\rm H}^2/\rho_0^2 \propto T^{-4}$, 
which is called the modified Kohler's rule
 \cite{Kimura}.
They were frequently cited as strong objections 
against a simple Fermi liquid picture, because analyses 
based on the relaxation time approximation (RTA) do not work.
However, recent theoretical works have revealed that 
the current vertex corrections (CVC's), 
which are dropped in the RTA, play important roles in HTSC's.
Due to the CVC's, anomalous behaviors of $R_{\rm H}$, $\Delta\rho/\rho$, 
thermoelectric power and Nernst coefficient are naturally explained 
{\it in a unified way, based on the Fermi liquid theory}
 \cite{Kontani-rev,Kontani-Hall,Kontani-MR,Kontani-S,Kontani-N}.

In the present paper,
we study the effect of a single nonmagnetic impurity on 
a Fermi liquid with strong AF fluctuations.
For that purpose, we developed a useful method of calculating 
the real space structure of the self-energy and the susceptibility
around the strong nonmagnetic impurity,
on the basis of an improved FLEX approximation.
When the AF fluctuations are strong,
we find that both local and staggered spin susceptibilities
are enhanced around the impurity.
Moreover, the residual resistivity $\Delta\rho$ per impurity,
which is determined by the nearly parallel shift of $\rho(T)$,
can take a huge value beyond the s-wave unitary scattering limit.
In addition, a ``Kondo-like'' insulating behavior ($d\rho/dT<0$)
emerges in the close vicinity of the AF-QCP.
These drastic impurity effects in nearly AF systems
come from the fact that the electronic states are modified 
in a wide range around the impurity, whose radius is 
approximately AF correlation length, $\xi_{\rm AF}$.
The present study provides a unified understanding for
various experimental impurity effect in HTSC's,
{\it as universal phenomena in nearly AF Fermi liquids}.
Any exotic mechanisms (breakdown of the Fermi liquid state)
need not to be assumed to explain them.

We find that the simple FLEX approximation does not reproduce 
reliable electronic states around the impurity.
For example, the spin susceptibility ${\hat \chi}^s$ 
is {\it reduced} around the impurity.
This failure comes from the fact that the feedback effect
on the vertex correction for ${\hat \chi}^s$ is neglected
in the FLEX approximation.
To overcome this difficulty,
we propose a modified version of the FLEX approximation,
which we call the $GV^I$-FLEX method.

In the present work,
we want to calculate the impurity effect
at sufficient low temperatures, say 50K.
For this purpose, we have to work on a large real-space
cluster with a impurity site, which should be at least
64$\times$64 to obtain the correct bulk electric state
at lower temperatures.
The FLEX approximation for such a large cluster is, however,
almost impossible to perform numerically.
Here, we invented the method to calculate the single impurity
effect in a large cluster, using the fact that the
the range of the modification of electronic states 
due to the impurity is at least $5\sim6$ except for in 
heavily under-doped systems.

\subsection{Previous Theoretical Studies}
Effect of a nonmagnetic impurity embedded in the Hubbard model
or in the $t$-$J$ model had been studied theoretically
by various methods.
The $t$-$J$ cluster model ($\sim20$ sites) 
with a single nonmagnetic impurity 
was studied using the exact diagonalization method
 \cite{Ziegler,Poilblanc}.
They found that the AF correlation is enhanced around the impurity.
The same result is realized in two-dimensional Heisenberg models
with a vacant site
because quantum fluctuations are reduced around the impurity
 \cite{Bulut89,Sandvik}.
The same mechanism will account for the enhancement of the 
AF correlation around an impurity in $t$-$J$ model.
Moreover, in the $t$-$J$ model, an effective long-range
impurity potential is induced due to the many-body effect.
The authors discussed that non s-wave scattering
given by the effective long-range potential could
give rise to a huge residual resistivity per impurity,
beyond the s-wave unitary scattering limit.
It is noteworthy that an extended Gutzwiller approximation 
was applied for the impurity problem in the $t$-$J$ model 
\cite{Ogata}.

Also, the Hubbard model with a single nonmagnetic impurity 
was studied using the random phase approximation (RPA) 
in refs. \cite{Bulut01,Bulut00,Ohashi}.
By assuming an ``extended impurity potential'',
they explained that the local susceptibility is enhanced 
in proportion to $\chi_{\rm Q}$ $[\propto T^{-1}]$ of the host,
reflecting the lack of translational invariance.
Similar analysis based on a phenomenological AF fluctuation model
was done \cite{Prelovsek}.
The staggered susceptibilities is also enhanced due to the 
change of the local DOS (Friedel oscillation) \cite{Fujimoto}.
However, the RPA could not explain the 
enhancement of local and staggered susceptibilities 
when the ($\delta$-functional) onsite impurity potential,
which corresponds to Zn or Li substitution in HTSC's, is assumed.
Therefore, results given by the RPA are not universal in that they
are very sensitive to the strength of extended impurity potential.

In the present work based on the $GV^I$-FLEX approximation, 
we show that $\delta$-functional onsite nonmagnetic impurity 
causes the enhancement of $\chi^s$ universally.
As a result, onsite impurities cause a huge residual resistivity
due to the nonlocal widespread change on the self-energy.
We will see that the $GV^I$-method gives a unified understanding
of the impurity problem in HTSC.

\subsection{Experimental Results}
Here, we introduce several experimental results in HTSC's 
and the related systems, which we focus on in the present work.
In later sections, we will discuss the origin of these
experimental facts, and show that they are qualitatively
well explained in a unified way {\it as the effect of nonmagnetic 
impurities or residual disorders}, on the basis of the $GV^I$-method.

\vspace{3mm}
\noindent
{\bf (a) Magnetic properties}: \ \
In optimally or under-doped HTSC's,
a nonmagnetic impurity replacing a Cu site causes a localized momentum.
A Curie like uniform spin susceptibility
is induced by dilute doping of Zn in YBa$_2$Cu$_3$O$_{7-x}$ (YBCO).
The Curie constant $C$ per Zn in under-doped compounds ($x\approx0.34$)
is much larger than that in optimally doped ones ($x\approx0$)
 \cite{Alloul99-2}.
They also reports an interesting relation $C\propto\Delta\rho$.
Curie-like susceptibility was observed in Al-doped
La$_{2-\delta}$Sr$_\delta$CuO$_4$ (LSCO)
 \cite{Ishida96}.
In Zn-doped YBCO compounds,
site-selective $^{89}$Y NMR measurements revealed that 
both the local spin susceptibility \cite{Alloul94,Alloul00}
and the staggered susceptibility \cite{Alloul00-2}
are prominently enhanced around the Zn-site,
within the radius of the AF correlation length $\xi_{\rm AF}$.
The same result was obtained by the 
$^7$Li Knight shift measurement in Li-doped YBCO compounds
 \cite{Alloul99}, and by the $^{63}$Cu NMR measurement
in Zn-doped YBCO compounds
 \cite{Jullien00}.
These NMR studies show that the impurities does not trap holes,
contrary to the suggestion by refs. \cite{Ziegler,Poilblanc}.
This fact means that the impurity-induced local moments
result from the change of the magnetic properties 
of itinerant electrons around the impurity sites.

\vspace{3mm}
\noindent
{\bf (b) Resistivity}: \ \
Fukuzumi et al. observed $\rho(T)$ in Zn-doped YBCO and LSCO 
for Zn concentration $n_{\rm imp}=0.02\sim 0.04$
 \cite{Uchida}.
In over-doped systems,
the residual resistivity $\Delta\rho(T)$ per impurity, which is 
determined from nearly parallel shift of $\rho(T)$ by impurities,
is consistent with the value for 2D electron gas;
$\rho_{\rm imp}^0=(4\hbar/e^2)n_{\rm imp}/n$.
However, $\Delta\rho \sim (4\hbar/e^2)n_{\rm imp}/|1-n| 
\gg\rho_{\rm imp}^0$ in under-doped systems.
This fact suggests that the scattering cross section of Zn 
anomalously increases in under-doped HTSC's, as if an effective 
radius of impurity potential grows due to many-body effect.
In addition, upturn of $\rho(T)$ ($d\rho/dT<0$) is observed 
below 50K in under-doped compounds.

Such a prominent enhancement of residual resistivity $\Delta\rho$
is also observed in HF compounds near the AF-QCP, which is realized
under a critical pressure $P_{\rm c}$
 \cite{Jaccard1,Jaccard2}.
Famous examples are CeCu$_5$Au ($P_{\rm c}\approx3.4$GPa)
 \cite{CeCuAu},
CeRhIn$_5$ ($P_{\rm c}\approx2$GPa) and CeCoIn$_5$ ($P_{\rm c}\approx0$GPa)
 \cite{Ce115,Ce115-2}.
Note that the enhancement of $\Delta\rho$ has nothing to do with
the increase of the renormalization factor $z$ ($\ll 1$)
by applied pressure, because $\Delta\rho$
is independent of $z$ in the Fermi liquid theory.
In addition, the residual resistivity of an organic superconductor
$\kappa$-(BEDT-TTF)$_4$Hg$_{2.89}$Br$_8$ ($T_{\rm c}\approx4$K),
which is close to the AF-QCP at ambient pressure,
decreases to be about 10\% of the original value by applying the pressure
 \cite{Taniguchi}.
Such a drastic reduction of $\Delta\rho$ cannot be attributed to 
the change of the DOS by pressure.

Ando et al. have measured the resistivity in HTSC's
under high magnetic field ($\sim$60T), which totally 
suppresses the superconductivity \cite{Ando1,Ando2}.
In LSCO, they found that the insulating behavior emerges under
the original $T_{\rm c}$ at ${\bf B}=0$.
This upturn of $\rho$ occurs when $k_{\rm F}l \sim 13 \gg 1$ 
in the $ab$-plane ($l$ being the mean free path),
so it has nothing to do with conventional localization in bad metals.
Also, It will be independent of the opening of the pseudo-gap 
because the pseudo-gap temperature $T^\ast$ is much higher.
Moreover, neither the weak localization or the Kondo effect 
due to magnetic impurities cannot be the origin of the upturn,
because the (negative) magnetoresistance in the insulating region
is independent of the field direction, and the insulating behavior
persists under very high magnetic field.

Sekitani et al. also found similar insulating behavior ($d\rho/dT<0$)
in under-doped electron-doped systems, M$_{2-\delta}$Ce$_\delta$CuO$_4$
(M=Nd, Pr, and La), in the normal state under high magnetic field
 \cite{Sekitani}.
They discussed that the residual apical oxygens (about 1\%),
which works as impurity scattering potentials, give rise to
the insulating behavior.
They expected that the Kondo effect occurs.
However,  the ${\bf B}$-dependence of $\rho({\bf B},T)$ does not
seem to be consistent with the Kondo effect.
The upturn of $\rho_c$ along the $c$-axis is also observed
in optimally doped Sm$_{2-\delta}$Ce$_\delta$CuO$_4$
($\delta\approx0.14$) in ref. \cite{Shibauchi};
the upturn is robust against the strong magnetic field ($\sim45$T),
both for ${\bf B}\parallel {\bf a}$ and ${\bf B}\parallel {\bf c}$.
Recently, upturn of $\rho$ in optimally doped PCCO had been observed
 \cite{Greene-upturn}.

\vspace{3mm}
\noindent
{\bf (c) Local density of states}: \\ 
STM measurement \cite{Pan}
revealed that a single nonmagnetic impurity 
in optimally doped Bi$_2$Sr$_2$CaCu$_2$O$_{8+\delta}$
causes strong suppression of the superconducting state
with radius $\sim15$\AA.
This ``Swiss cheese structure'' below $T_{\rm c}$ had been suggested 
by the $\mu$-SR measurement \cite{Uemura} as well as the 
specific heat measurement in Zn doped LSCO \cite{Ido2}.
Reference \cite{Ido2} reports that the radius of Swiss cheese hole 
increases as the carrier doping decreases.
The radius is approximately $\xi_{\rm AF}$, 
rather than the coherence length.
This result suggests that the electronic properties 
in the Swiss cheese hole is strongly modified even above $T_{\rm c}$,
because $\xi_{\rm AF}$ is a characteristic length scale
in the normal state.
Moreover, resent STM/STS measurements have revealed that the local 
density of states (DOS) in Bi$_2$Sr$_2$CaCu$_2$O$_{8+x}$
is very inhomogeneous in the atomic scale
 \cite{Davis,Ido}.
The observed ununiformity originates from the weak scattering
potentials from out-of-plane dopant atoms
 \cite{Davis}.
These experimental observations suggest that the
DOS and the electronic states in HTSC's are quite
sensitive to the disorder potential.

\section{Formalism}
\label{sec:formalism}
In the present paper, we study a $(N\times N)$ square lattice
Hubbard model with an impurity site:
\begin{eqnarray}
H&=& H_0+H_{\rm imp} ,
  \label{eqn:Ham} 
 \\
H_0&=& \sum_{\k\s}\e_\k c_{\k\s}^\dagger c_{\k\s}
+ U\sum_i n_{i \uparrow} n_{i \downarrow} ,
 \\
H_{\rm imp}&=& I (n_{0 \uparrow} + n_{0 \downarrow}) ,
 \label{eqn:Himp}
\end{eqnarray}
where $H_0$ is the Hubbard model for the host system.
In $H_{\rm imp}$, $I$ is the onsite nonmagnetic impurity potential 
at the origin (${\bf r}=0$).
Because the translational invariance is violated
in the case of $I\ne0$,
the self-energy $\Sigma({\bf r},{\bf r}'; \e_n)$ and 
the Green function $G({\bf r},{\bf r}'; \e_n)$
cannot be functions of ${\bf r}-{\bf r}'$.
In the present paper, we concentrate on the strong impurity 
potential case (unitary scattering case); $I=\infty$.
We will take the $I=\infty$-limit in the course of the calculation.

We develop the method to study the impurity effect 
with strong potential on the basis of the FLEX approximation.
When $I\ne0$, Green functions which compose the self-energy
for $I=0$ get insertions of $I$'s in all possible manners.
Then, the self-energy is divided into two terms:
\begin{eqnarray}
\Sigma({\bf r}_i,{\bf r}_j;\e_n)= 
\Sigma^{0}({\bf r}_i-{\bf r}_j;\e_n)+ \delta\Sigma({\bf r}_i,{\bf r}_j;\e_n),
\end{eqnarray}
where $\Sigma^{0}$ is the self-energy for $I=0$, that is,
the self-energy for the host system without impurity.
$\delta\Sigma$ represents the cross terms between $I$ and $U$.
Here, $\delta\Sigma$ does not contain any terms composed of only $I$'s.
The full Green function for $H_0+H_{\rm imp}$ is composed of
$\Sigma^0$, $\delta\Sigma$ and $I$.
In the case of $I=\infty$,
$\delta\Sigma({\bf r},0)=-\delta\Sigma(0,{\bf r})=-\Sigma^0({\bf r},0)$
because $\Sigma({\bf r},0)=\Sigma(0,{\bf r})=0$.
The impurity potential also causes the
nonlocal change in the self-energy, that is,
$\delta\Sigma({\bf r}_i,{\bf r}_j)$ is finite 
even for ${\bf r}_i, {\bf r}_j \ne 0$.
$\delta\Sigma({\bf r}_i,{\bf r}_j)$ 
will quickly converge to zero as ${\bf r}_i$ or ${\bf r}_j$
are away from the origin.

The Dyson equation for the real-space 
Green function in the matrix representation is given by
\begin{eqnarray}
{\hat G}(\e_n)
&=& {\hat G^{00}}(\e_n)
+ {\hat G^{00}}(\e_n) ( {\hat \Sigma}(\e_n)+{\hat I} ) {\hat G}(\e_n) 
 \nonumber \\
&=& {\hat G^{0}}(\e_n)
+ {\hat G^{0}}(\e_n) (\delta{\hat \Sigma}(\e_n)+{\hat I} ) {\hat G}(\e_n) ,
\end{eqnarray}
where $({\hat I})_{i,j}= I\delta_{i,0}\delta_{j,0}$
represents the impurity potential at the origin.
$G^{00}({\bf r}_i-{\bf r}_j; \e_n)
=\frac1{N^2}\sum_\k (i\e_n+\mu-\e_\k)^{-1} e^{i\k \cdot ({\bf r}_i-{\bf r}_j)}$ 
is the non-interacting Green function, and
${\hat G^{0}}(\e_n)=([{\hat G^{00}}(\e_n)]^{-1}
 -{\hat \Sigma^0}(\e_n))^{-1}$ 
is the interacting Green function without impurity ($I=0$).

\begin{figure}
\begin{center}
\epsfig{file=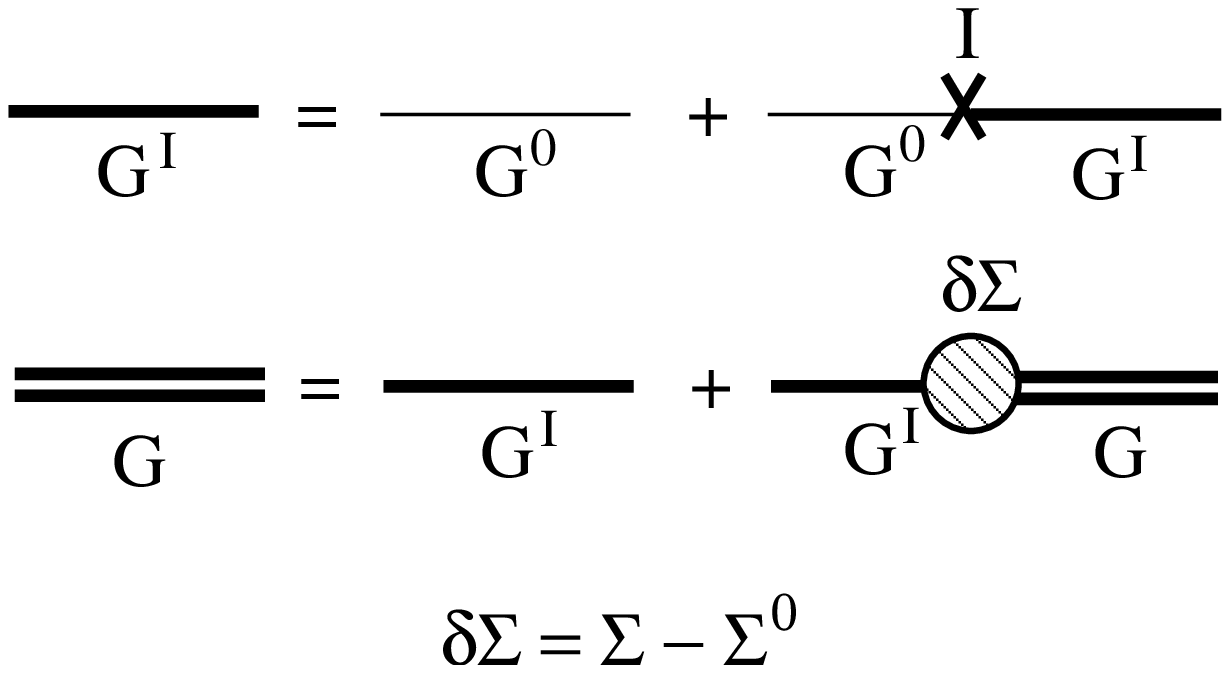,width=6cm}
\end{center}
\caption{Dyson equations for $G^I$ and $G$ 
in the presence of $I$ and $\delta\Sigma$.
}
\label{fig:Dyson}
\end{figure}

The Dyson equation is also written as
\begin{eqnarray}
{\hat G}(\e_n)
&=& {\hat G}^{I}(\e_n)
+ {\hat G}^{I}(\e_n) \delta{\hat \Sigma}(\e_n) {\hat G}(\e_n) ,
 \label{eqn:Dyson-I} \\
{\hat G}^{I}(\e_n)
&=& {\hat G^{0}}(\e_n)+ {\hat G^{0}}(\e_n){\hat I}{\hat G}^{I}(\e_n) ,
 \label{eqn:GI}
\end{eqnarray}
where ${\hat G}^{I}$ represents the Green function
composed of $I$ and $G^0$, that is,
${\hat G}^{I}={\hat G}|_{\delta\Sigma=0}$.
Equations (\ref{eqn:Dyson-I}) and (\ref{eqn:GI})
are expressed in Fig.\ref{fig:Dyson}.

In the present work, we calculate 
the self-energy for $I\ne0$ based on the improved FLEX approximation,
without taking any averaging with respect to the position of impurity.
Hereafter, we propose the three versions of 
frameworks for calculating the self-energy as follows.

\vspace{3mm}
\noindent
{\bf (I) $GV^0$-method} : \
First, we introduce the ``$GV^0$-FLEX approximation'',
where the full Green function $G$ is obtained self-consistently
whereas any impurity effects on the effective interaction $V^0$
are neglected.
Here, the self-energy is given by
\begin{eqnarray}
\Sigma_{\rm [GV0]}({\bf r}_i,{\bf r}_j;\e_n)
&=& T\sum_l G_{\rm [GV0]}({\bf r}_i,{\bf r}_j;\w_l+\e_n) 
 \nonumber \\
& &\times V^0({\bf r}_i,{\bf r}_j;\w_l) ,
 \label{eqn:self-GV0} 
\end{eqnarray}
where $\e_n= (2n+1)\pi T$ and $\w_l= 2l\cdot\pi T$, respectively.
$G_{\rm [GV0]}$ and 
$\delta\Sigma_{\rm [GV0]} \equiv \Sigma_{\rm [GV0]}-\Sigma^0$
satisfy the Dyson equation, (\ref{eqn:Dyson-I}).
$V^0$ and $\Sigma^0$ are given by the FLEX approximation for the host system.
Hereafter, we call eq. (\ref{eqn:self-GV0}) the $GV^0$-method for simplicity,
because the self-energy is symbolically written as $G\circ V^0$. 
Here, $V^0$ and $\Sigma^0$ are given by
\begin{eqnarray}
{V}^0({\bf r}_i,{\bf r}_j;\w_l)
&=& \frac1{N^2}\sum_\q {V}^0(\q,\w_l)e^{i\q\cdot({\bf r}_i-{\bf r}_j)},
 \\
V^0(\q,\w_l)
&=& U^2 \left( \frac32 {\chi}_\q^{0s}(\w_l) 
  +\frac12 {\chi}_\q^{0c}(\w_l) 
  - {\Pi}_\q^0(\w_l) \right) \mbox{,}
    \nonumber \\ \\
\chi_\q^{0s(c)}(\w_l)
 &=& {\Pi}_\q^0(\w_l) \cdot \left\{ {1} -(+)
  U{\Pi}_\q^0(\w_l) \right\}^{-1} \mbox{,} \\
\Pi_\q^0(\w_l)
 &=& -T\sum_{\k, n} G_{\q+\k}^0(\w_l+\e_n) G_\k^0(\e_n) \mbox{,}
 \\
\Sigma_\k^0(\e_n)&=& T\sum_{\q,l}G_{\k+\q}^0(\e_n+\w_l)V^0(\q,\w_l) ,
\end{eqnarray}
where $G_\k^0(\e_n)=(i\e_n+\mu-\e_k-\Sigma_\k^0(\e_n))^{-1}$.
$\chi_\q^{0s}$ and $\chi_\q^{0c}$ are 
spin and charge susceptibilities, respectively,
given by the FLEX approximation for the host system.

Using the $GV^0$-method,
We can calculate the nonlocal change in the self-energy
induced around the impurity, $\delta\Sigma$.
However, the nonlocal change in the spin susceptibility
is not taken into account in the $GV^0$-method.
To calculate this effect, we introduce other two methods
as follows.

\vspace{3mm}
\noindent
{\bf (II) $GV$-method} : \
Next, we explain the ``$GV$-method'',
which is equal to the FLEX approximation in real space.
In this method, both the self-energy and the effective interaction
are obtained fully self-consistently.
The self-energy in the $GV$-method is given by
\begin{eqnarray}
\Sigma_{\rm [GV]}({\bf r}_i,{\bf r}_j;\e_n)
&=& T\sum_l G_{\rm [GV]}({\bf r}_i,{\bf r}_j;\w_l+\e_n) ,
 \nonumber \\
& &\times V({\bf r}_i,{\bf r}_j;\w_l)
 \label{eqn:self-GV} \\
{\hat V}(\w_l)
&=& U^2\left( \frac32{\hat \chi}^s(\w_l)
 + \frac12{\hat \chi}^c(\w_l) -{\hat \Pi}(\w_l) \right) ,
 \nonumber \\
 \label{eqn:self-V}
\end{eqnarray}
where $G_{\rm [GV]}$ and 
$\delta\Sigma_{\rm [GV]}=\Sigma_{\rm [GV]}-\Sigma^0$ satisfy
Dyson equation (\ref{eqn:Dyson-I}).
The spin and charge susceptibilities in the $GV$-method,
${\hat \chi}^{s}$ and ${\hat \chi}^{c}$, are given by 
\begin{eqnarray}
{\hat \chi}^{s(c)}
&=& {\hat \Pi} \left( 1 -(+) U{\hat \Pi} \right)^{-1} ,
 \label{eqn:chisc-GV}
 \\
{\Pi}({\bf r}_i,{\bf r}_j;\w_l)
&=& -T\sum_{\e_n} G_{\rm [GV]}({\bf r}_i,{\bf r}_j;\e_n+\w_l)
 \nonumber \\
& &\times G_{\rm [GV]}({\bf r}_j,{\bf r}_i;\e_n) .
 \label{eqn:Pi-GV}
\end{eqnarray}
In the $GV$-method,
the impurity effect on $V$ is fully taken into account
in terms of the FLEX approximation.
However, we find that the numerical results given by the 
$GV$-method are totally inconsistent with experimental facts.
This is because the vertex corrections (VC's) 
for the spin susceptibility,
which are dropped in the $GV$-method,
becomes significant in strongly correlated systems.
We will discuss this point in \S \ref{sec:VC}.

\vspace{3mm}
\noindent
{\bf (III) $GV^I$-method}: \
To overcome the difficulty inherent in the $GV$-method,
we propose the ``$GV^I$-method'', where
the Green function is obtained self-consistently
whereas the impurity effect on the effective interaction
is calculated in a partially self-consistent way.
We will show that the $GV^I$-method is the most superior among (I)-(III).
Here, the self-energy is given by
\begin{eqnarray}
\Sigma_{\rm [GVI]}({\bf r}_i,{\bf r}_j;\e_n)
&=& T\sum_l G_{\rm [GVI]}({\bf r}_i,{\bf r}_j;\w_l+\e_n) 
 \nonumber \\
& &\times V^I({\bf r}_i,{\bf r}_j;\w_l) ,
 \label{eqn:self-GVI} \\
{\hat V}^I(\w_l)
&=& U^2\left( \frac32{\hat \chi}^{Is}(\w_l)
 + \frac12{\hat \chi}^{Ic}(\w_l) -{\hat \Pi}^I(\w_l) \right) ,
 \nonumber \\
 \label{eqn:self-VI}
\end{eqnarray}
where $G_{\rm [GVI]}$ and 
$\delta\Sigma_{\rm [GVI]}=\Sigma_{\rm [GVI]}-\Sigma^0$ satisfy
Dyson equation (\ref{eqn:Dyson-I}).
The spin and charge susceptibilities in the $GV^I$-method,
${\hat \chi}^{Is}$ and ${\hat \chi}^{Ic}$, are given by 
\begin{eqnarray}
{\hat \chi}^{Is(c)}
&=& {\hat \Pi} \left( 1 -(+) U{\hat \Pi^I} \right)^{-1} ,
 \label{eqn:chisc-GVI} \\
{\Pi}^I({\bf r}_i,{\bf r}_j;\w_l)
&=& -T\sum_{\e_n} G^I({\bf r}_i,{\bf r}_j;\e_n+\w_l)
 \nonumber \\
& &\times G^I({\bf r}_j,{\bf r}_i;\e_n) .
 \label{eqn:Pi-GVI}
\end{eqnarray}
In ${\hat \chi}^{Is}$ and ${\hat \chi}^{Ic}$ in the $GV^I$-method,
the self-energy correction given by the 
cross terms between $I$ and $U$ [$\delta\Sigma$]
is not taken into account, whereas it is taken in the $GV$-method.
Nonetheless, results given by the $GV^I$-method are
completely different from results by the $GV$-method,
and the former results are well consistent with experiments.
For example, ${\hat \chi}^{Is}$ given in 
eq.(\ref{eqn:chisc-GVI}) is strongly enhanced
around the impurity, whereas ${\hat \chi}^{s}$ given in 
eq.(\ref{eqn:chisc-GV}) is suppressed by $\delta\Sigma$.
In \S \ref{sec:VC}, we will show that the latter result is an 
artifact of the $GV$-method because the reduction of ${\hat \chi}^{s}$
due to $\delta\Sigma$ is overestimated.
As a result, the $GV^I$-method is much superior to the $GV$-method.

\begin{figure}
\begin{center}
\epsfig{file=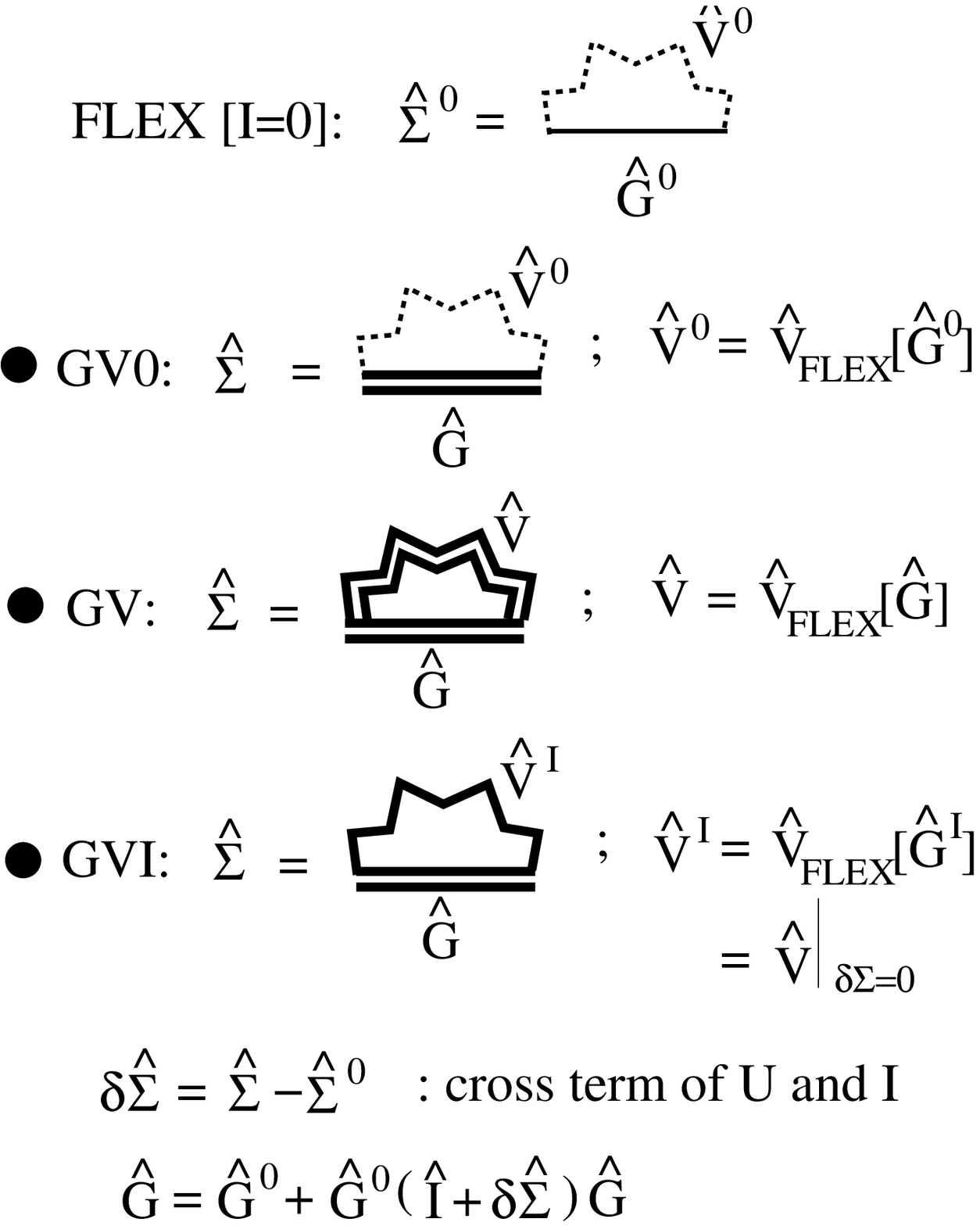,width=7cm}
\end{center}
\caption{Self-energy for the Hubbard model with a
nonmagnetic impurity, given by the $GV$, $GV^I$ and $GV^0$-method.
}
  \label{fig:Self}
\end{figure}

Figure \ref{fig:Self} expresses the 
self-energies for $GV$, $GV^I$ and $GV^0$-methods, respectively.
These methods are equivalent to the FLEX approximation when $I=0$.
In later sections, 
we solve the single impurity problem in the presence of 
the Coulomb interaction on the basis of these three methods.

\section{Method of Numerical Calculation}
\label{sec:method}

\begin{figure}
\begin{center}
\epsfig{file=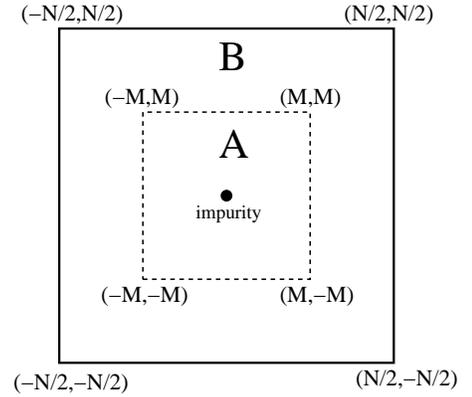,width=6cm}
\end{center}
\caption{Schematic expression for region A and B.
}
  \label{fig:region}
\end{figure}

In this section, we study the two-dimensional
Hubbard model with an impurity potential at the origin.
We work on the $(N\times N)$-square lattice with 
periodic boundary condition.
$N$ should be large (at least 64) enough to achieve the 
thermodynamic limit at low temperatures.
On the other hand, the range of nonlocal change in 
electronic states due to the impurity is only a few lattice spacings.
Taking this fact into account, 
we calculate $\delta\Sigma_{\a,\b}$ ($\a$ and $\b$
being the lattice points in real space) by the (improved) FLEX
approximation only for $|\a|, |\b| \le M$ ($M\ll N$), and we put 
$\delta\Sigma_{\a,\b}=0$ for $|\a| > M$ or $|\b| > M$.
Here we put $M=6\sim8$, which is enough to obtain 
a reliable numerical result.
Here, we explain the method how to
reduce the working area to a $((2M+1)\times (2M+1))$-square lattice 
in solving the single impurity problem in the $(N\times N)$-square lattice.
This technique helps us to avoid considerable numerical difficulty.
Hereafter, we use $i,j,\cdots$ ($\a,\b,\cdots$) to represent 
the lattice points in the region A+B (region A) 
in Fig.\ref{fig:region}.

According to eq.(\ref{eqn:GI}),
$G^I_{i,j}$ is given by
\begin{eqnarray}
G^I_{0,j}&=& G^I_{j,0}= \frac{G^0_{0,j}}{1-I G^0_{0,0}},
 \label{eqn:GI0} \\
G^I_{i,j}&=& G^0_{i,j} + I \frac{G^0_{i,0}G^0_{0,j}}{1-I G^0_{0,0}}
 \ \ \ \mbox{for $i,j\ne0$},
 \label{eqn:GIij}
\end{eqnarray}
where $G^0_{i,j}(\e_n)=G^0({\bf r}_i,{\bf r}_j;\e_n)= 
 G^0({\bf r}_i-{\bf r}_j;\e_n)= \frac1{N^2} \sum_\k 
 G^0_\k(\e_n)\exp(i\k\cdot({\bf r}_i-{\bf r}_j)$;
$G^0_\k(\e_n)$ is the Green function given by the FLEX
approximation without impurity potential.
Therefore, $G^I_{i,j}$ in the limit of $I=\infty$ is 
\begin{eqnarray}
G^I_{0,j}&=& G^I_{j,0}= 0 ,
 \label{eqn:GI0j-Iinf} \\
G^I_{i,j}&=& G^0_{i,j} - \frac{G^0_{i,0}G^0_{0,j}}{G^0_{0,0}} 
 \ \ \ \mbox{for $i,j\ne0$} .
 \label{eqn:GIij-Iinf}
\end{eqnarray}

According to eq.(\ref{eqn:Dyson-I}),
the Dyson equation for the full Green function 
$G_{\a,\b}$ inside of region A is given by 
\begin{eqnarray}
G_{\a,\b}&=& G^I_{\a,\b}+ \sum_{\b'}^{\rm A} A_{\a,\b'}G_{\b',\b} ,
 \label{eqn:Dyson} \\
A_{\a,\b}&=& \sum_{\b'}^{\rm A} G^I_{\a,\b'}\delta\Sigma_{\b',\b} .
\end{eqnarray}
Note that $\delta\Sigma_{\a,\b}$ is finite only when
both $\a$ and $\b$ are inside of region A
in the present approximation.
By solving eq.(\ref{eqn:Dyson}), we obtain
\begin{eqnarray}
{\hat G}(\e_n)&=& \left( {\hat 1}-{\hat A}(\e_n) \right)^{-1} 
 {\hat G}^I(\e_n) ,
\end{eqnarray}
which is a ($(2M+1)^2\times (2M+1)^2$)-matrix equation.
In the limit of $I=\infty$, $G_{\a,\b}$ is given by
\begin{eqnarray}
G_{\a,\b} &=&
 \sum_{\delta\ne0}^{\rm A} \left( {\hat 1}-{\hat A} \right)_{\a\delta}^{-1}
 \left( G^0_{\delta,\b}-\frac{G^0_{\delta,0}G^0_{0,\b}}{G^0_{0,0}}
 \right) 
 \label{eqn:GIab-Iinf}
\end{eqnarray}
Note that $G_{\a,0}=G_{0,\b}=0$.
$A_{\a,\b}$ in eq.(\ref{eqn:GIab-Iinf}) is given by
\begin{eqnarray}
A_{\a,\b}&=& D_{\a\b}^{0}-\frac{G^0_{\a,0}}{G^0_{0,0}}D_{0,\b}^{0},
 \\
D_{\a,\b}^{0}&=& \sum_{\b'}^{\rm A} G_{\a,\b'}^0\delta\Sigma_{\b',\b} .
 \label{eqn:D0}
\end{eqnarray}
%

Next, we study the spin and charge susceptibilities
in the presence of an impurity.
In the FLEX approximation,
the equation for the magnetic susceptibility in real space is
\begin{eqnarray}
\chi_{i,j}^s = \Pi_{i,j}+ U \sum_{l}^{\rm A+B} \Pi_{i,l}\chi_{l,j}^s ,
 \label{eqn:chis-org}
\end{eqnarray}
where $i$, $j$ and $l$ represent the lattice points in region A+B.
$\Pi_{i,j}$ is the irreducible susceptibility 
defined in eq.(\ref{eqn:Pi-GV}) in the $GV$-method
or in eq.(\ref{eqn:Pi-GVI}) in the $GV^I$-method.
This equation is not easy to solve because of its huge
matrix size; $(N^2\times N^2)$.
To solve this difficulty, we reduce the above equation
to $((2M+1)^2\times (2M+1)^2)$-matrix equation (inside of region A),
approximating that $\Pi_{i,j}=\Pi_{i-j}^0$
when $i$ and/or $j$ are in region B.
Here, $\Pi_{i-j}^0$ is the irreducible susceptibility
for the host system:
$\Pi_{i-j}^0(\w_l)= -T\sum_{\e_n}G_{i-j}^0(\e_n+\w_l)G_{j-i}^0(\e_n)$.

Here, $\chi_{\a,\b}^s$ inside region A is rewritten as
\begin{eqnarray}
\chi_{\a,\b}^s &=& 
 \Pi_{\a,\b}+U\sum_{\gamma\ne0}^{\rm A}\Pi_{\a,\gamma}\chi_{\gamma,\b}^s
 \nonumber \\
& &+ {\delta \chi}_{\a,\b}^s+U\sum_{\gamma\ne0}^{\rm A}
 {\delta \chi}_{\a,\gamma}^s\chi_{\gamma,\b}^s,
 \label{eqn:chis-I} \\
{\delta \chi}_{\a,\b}^s &=&
 U\sum_l^{\rm B} \Pi_{\a,l}^0\Pi_{l,\b}^0
 \nonumber \\
& &+U^2\sum_{l,m}^{\rm B} \Pi_{\a,l}^0\Pi_{l,m}^0\Pi_{m,\b}^0
 +\cdots .
 \label{eqn:chis-bar}
\end{eqnarray}
The infinite series of summation in eq.(\ref{eqn:chis-bar})
can be taken by using the following fact: In the case of $I=0$,
eq.(\ref{eqn:chis-I})
gives $\chi_{\a-\b}^{0s} \equiv\frac1{N^2}\sum_\k 
 \Pi_\k^{0}(1-U\Pi_\k^0)^{-1} \exp(i\k\cdot({\bf r}_\a-{\bf r}_\b))$.
As a result, we obtain that
\begin{eqnarray}
\chi_{\a,\b}^{0s} &=& 
 \Pi_{\a,\b}^0+U\sum_{\gamma}^{\rm A}\Pi_{\a,\gamma}^0\chi_{\gamma,\b}^{0s}
 \nonumber \\
& &+ {\delta \chi}_{\a,\b}^{s}+U\sum_{\gamma}^{\rm A}
 {\delta \chi}_{\a,\gamma}^s\chi_{\gamma,\b}^{0s},
 \label{eqn:chis0-I} \\
\end{eqnarray}
where the summation of $\gamma$ contains the origin.
By solving this $((2M+1)^2\times (2M+1)^2)$-matrix equation, 
${\delta \chi}^{s}$ is obtained as
\begin{eqnarray}
{\hat {\delta \chi}}^{s}
 = \left( {\hat \chi}^{0s}-{\hat \Pi}^{0}
  -U{\hat \Pi}^{0}{\hat \chi}^{0s} \right) 
 \left( 1+U{\hat \chi}^{0s} \right)^{-1} .
 \label{eqn:chis-bar2}
\end{eqnarray}
Using eq.(\ref{eqn:chis-bar2}),
the solution of eq.(\ref{eqn:chis-I}) is given by
\begin{eqnarray}
{\hat \chi}^s&=& \left( 1-U{\hat \Pi} -U{\hat {\delta \chi}}^{s*}
 \right)^{-1} \left( {\hat \Pi}+{\hat {\delta \chi}}^{s*} \right) ,
\label{eqn:chis-I2} \\
{\delta \chi}_{\a,\b}^{s*}&\equiv& 
 {\delta \chi}_{\a,\b}^{s} (1-\delta_{\a\b,0}) ,
 \label{eqn:deltachis_ast}
\end{eqnarray}
where the factor $1-\delta_{\a\b,0}$ in eq. (\ref{eqn:deltachis_ast})
represents the elimination of $\gamma=0$ in the summation in
eq. (\ref{eqn:chis-I}).
Numerical calculation of eq.(\ref{eqn:chis-I2}) is easy because 
its matrix size $((2M+1)^2\times (2M+1)^2)$ is not large
for $M=6\sim8$.
In the numerical study, we have to check that all the eigenvalues of 
$1-U{\hat \Pi} -U{\hat {\delta \chi}}^{s*}$ in eq.(\ref{eqn:chis-I2}) 
are positive, because a single impurity in a paramagnetic bulk system
could not induce a static magnetic order.

In the same way, we derive the charge susceptibility, 
which is given by the following $(N^2\times N^2)$-matrix equation:
\begin{eqnarray}
\chi_{i,j}^c = \Pi_{i,j}- U \sum_{l}^{\rm A+B} \Pi_{i,l}\chi_{l,j}^c .
 \label{eqn:chic-org}
\end{eqnarray}
Taking the same procedure as used in deriving eq.(\ref{eqn:chis-I2}),
$\chi_{\a,\b}^c$ in the region A is given by the following
$((2M+1)^2\times (2M+1)^2)$-matrix:
\begin{eqnarray}
{\hat \chi}^c&=& \left( 1+U{\hat \Pi} +U{\hat {\delta \chi}}^{c*}
 \right)^{-1} \left( {\hat \Pi}+{\hat {\delta \chi}}^{c*} \right) .
\label{eqn:chic-I2} 
\end{eqnarray}
Here, ${\delta \chi}_{\a,\b}^{c*}$ is given by
\begin{eqnarray}
{\delta \chi}_{\a,\b}^{c*}&\equiv& 
 {\delta \chi}_{\a,\b}^{c} (1-\delta_{\a\b,0}), \\
{\hat {\delta \chi}}^{c} 
 &=& \left( {\hat \chi}^{0c}-{\hat \Pi}^{0}
  +U{\hat \Pi}^{0}{\hat \chi}^{0c} \right) 
 \left( 1-U{\hat \chi}^{0c} \right)^{-1} ,
 \label{eqn:chic-bar2} 
\end{eqnarray}
where ${\chi}_{i-j}^{0c}=\frac1{N^2}\sum_\k 
 \Pi_\k^{0}(1+U\Pi_\k^0)^{-1} \exp(i\k\cdot({\bf r}_i-{\bf r}_j))$
is the charge susceptibility for the host system.

Finally, the spin and charge susceptibilities in the 
$GV^I$-method, ${\hat \chi}^{Is}$ and ${\hat \chi}^{Ic}$, 
are obtained as follows:
\begin{eqnarray}
{\hat \chi}^{Is}&=& \left( 1-U{\hat \Pi}^I -U{\hat {\delta \chi}}^{s*}
 \right)^{-1} \left( {\hat \Pi}^I+{\hat {\delta \chi}}^{s*} \right),
\label{eqn:chis-I3} \\
{\hat \chi}^{Ic}&=& \left( 1+U{\hat \Pi}^I +U{\hat {\delta \chi}}^{c*}
 \right)^{-1} \left( {\hat \Pi}^I+{\hat {\delta \chi}}^{c*} \right) ,
\label{eqn:chic-I3} 
\end{eqnarray}
where ${\hat \Pi}^I$ is given in eq.(\ref{eqn:Pi-GVI}).
Note that we have to check that all the eigenvalues of 
$1-U{\hat \Pi}^I -U{\hat {\delta \chi}}^{s*}$ in eq.(\ref{eqn:chis-I3}) 
are positive in the numerical study.

\section{Numerical Results for Local DOS and Spin Susceptibilities}
\label{sec:numerical}
In this section, we show several numerical results 
given by $GV^0$, $GV^I$ and $GV$ methods.
In each methods, the self-energy is obtained self-consistently.
We find that $GV^I$-method gives the most reliable results,
irrespective that the fully self-consistent condition 
for the quasiparticle interaction is not imposed.
In the $GV$-method, on the other hand,
the reduction of $\chi^s$
due to $\delta\Sigma$ [i.e., the nonlocal change in the self-energy 
given by the cross terms between $I$ and $U$] is overestimated.
In \S \ref{sec:VC}, we will show that the reduction of $V$ due to 
$\delta\Sigma$ is almost recovered if one takes account of the VC 
due to the excess spin fluctuations induced by the impurity.
By this reason, the $GV^I$-method is the most reliable formalism.
In the present section, we mainly focus on the numerical results 
given by the $GV^I$-method.

In the present numerical study, 
the dispersion of the conduction electron is given by
\begin{eqnarray}
\e_\k&=& 2t(\cos(k_x)+\cos(k_y))
 + 4t'\cos(k_x)\cos(k_y)
 \nonumber \\
& & + 2t''(\cos(2k_x)+\cos(2k_y)),
\end{eqnarray}
where $t$, $t'$, and $t''$ are the nearest, the next nearest, 
and the third nearest neighbor hopping integrals, respectively.
In the present study, we use the following set of parameters:
(I) YBCO (hole-doped):
 $t_0=-1$, $t_1=1/6$, $t_2=-1/5$, $U=6\sim8$.
(II) NCCO (electron-doped):
 $t_0=-1$, $t_1=1/6$, $t_2=-1/5$, $U=5.5$.
(III) LSCO (hole-doped):
  $t_0=-1$, $t_1=1/10$, $t_2=-1/10$, $U=4\sim5$.
These hopping parameters are equal to those
used in ref.\cite{Kontani-Hall}.
They were determined qualitatively by fitting to the Fermi surface 
(FS) given by ARPES measurements or the LDA band calculations.
The shape of the FS's for (I)-(III), all of which are hole-like,
are shown in ref.
 \cite{Kontani-Hall}.
Because $t_0\sim4000$K in real systems, $T=0.01$ corresponds to $\sim40$K.
In the present numerical study,
$64\times64$ $\k$-meshes and 1024 Matsubara frequencies are used.

The value of $U$ used in the present study is rather smaller 
than the bandwitdh ($W_{\rm band}$), irrespective that
the real coulomb interaction is larger than $W_{\rm band}$.
This will be justified by considering that $U$ used here
is the {\it effective} Coulomb interaction $U_{\rm eff}$
between quasiparticles with low energies.
In fact, $U_{\rm eff} \sim W_{\rm band}$ according to
the Kanamori theory based on the two-particle approximation
 \cite{Kanamori}.

In Ref.\cite{Kontani-Hall},
the shape of the Fermi surface, the temperature and
the momentum dependences of the spin susceptibility [$\chi_\q^s(0)$]
and the quasiparticle damping rate [Im$\Sigma_\q(-i\delta)$]
given by the FLEX approximation are explained in detail.
The obtained results are well consistent with experiments.
For example, the $\q$-dependence of $\chi_\q^s(0)$ shows that
$\xi_{\rm AF}=2\sim3$a in YBCO (n=0.85) at $T=0.02$,
which is consistent with neutron measurements.

\subsection{Local Density of States}
\label{sec:DOS}

\begin{figure}
\begin{center}
\epsfig{file=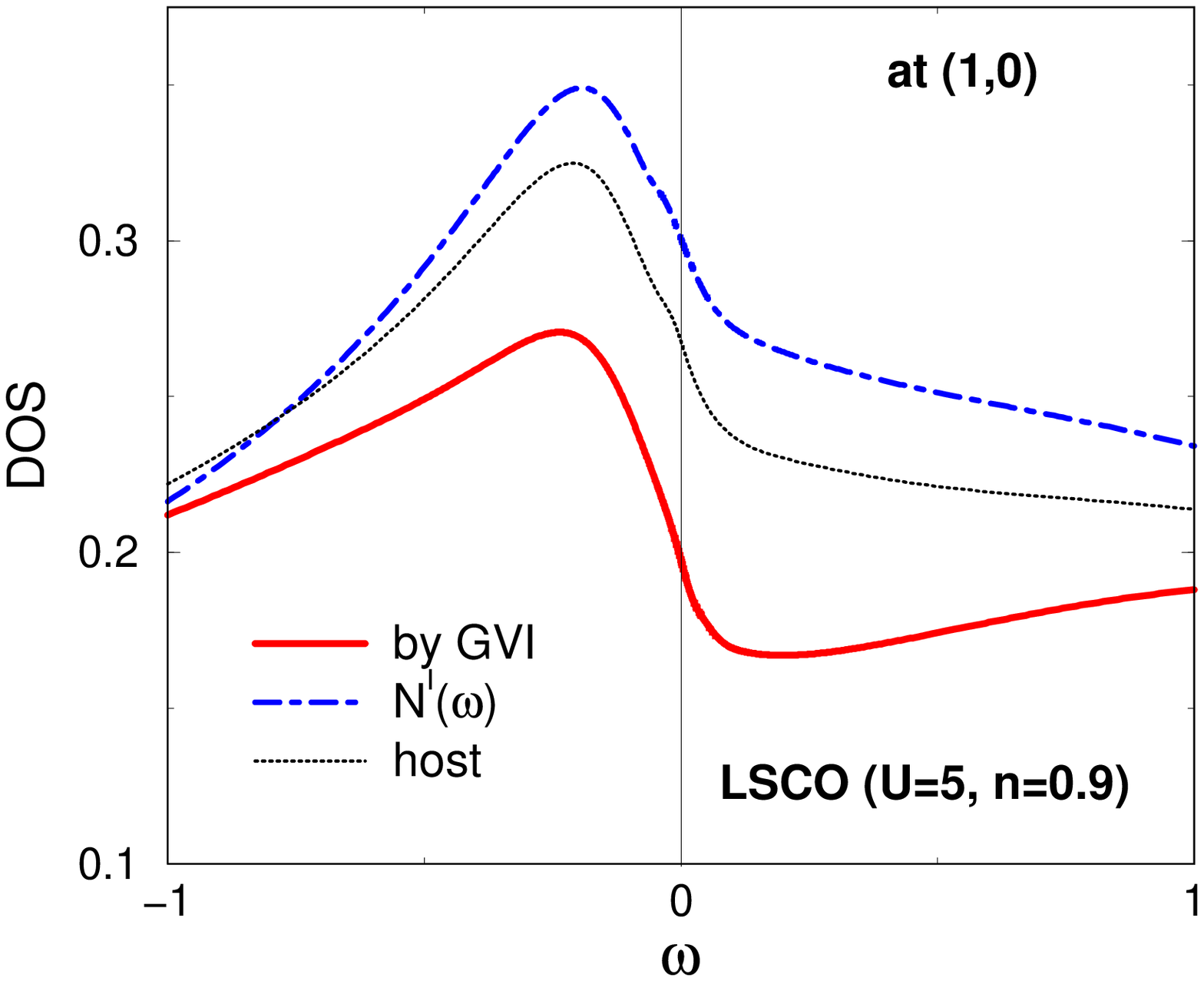,width=7cm}
\epsfig{file=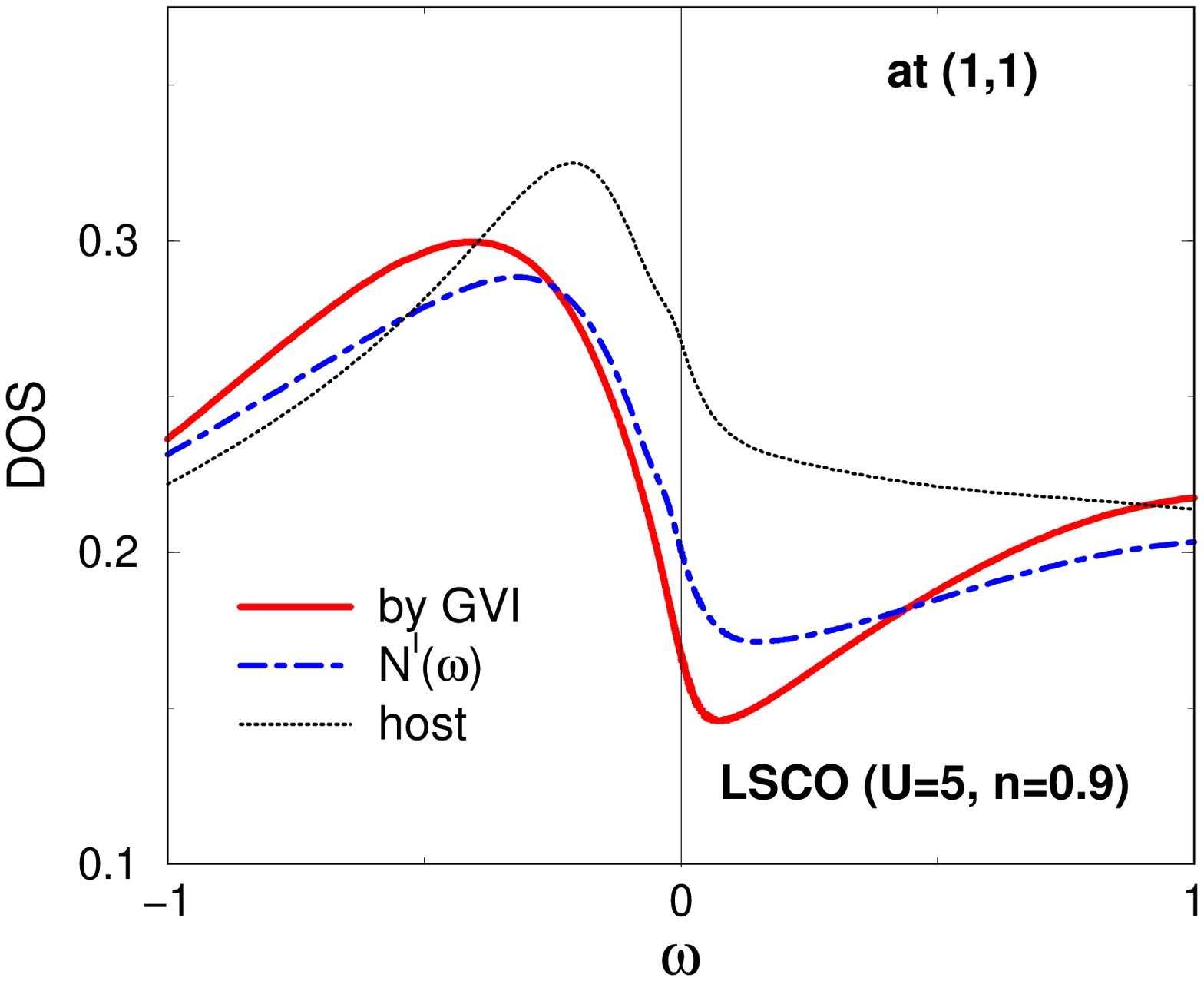,width=7cm}
\end{center}
\caption{(Color online) 
Obtained local DOS for LSCO ($n=0.9$, $U=5$) at $T=0.02$ ($\sim80$K).
(i)host: DOS given by the FLEX approximation without impurity.
(ii) $N_l^I(\w)=-\frac1{N^2}\sum_\k {\rm Im}G^I(\w)$
where $G^I$ is given in eq.(\ref{eqn:GIij-Iinf}).
(iii) local DOS given by the $GV^I$-method;
$-\frac1{N^2}\sum_\k {\rm Im}G_{\rm [GVI]}(\w)$
}
\label{fig:DOS}
\end{figure}

Figure \ref{fig:DOS} represents the density of states (DOS)
for a hole-doped system (LSCO, $n=0.9$) at $T=0.02$,
at the nearest-neighbor site (${\bf r}=(1,0)$) 
and the next-nearest-neighbor site (${\bf r}=(1,1)$)
of the impurity, respectively.
Here, ``host'' represents the DOS without the impurity
given by the FLEX approximation, 
$N_{\rm host}(\w)\equiv \frac1{\pi N^2}\sum_\k {\rm Im}G_\k^0(\w-i\delta)$.
On the other hand, 
$N_l^I(\w)\equiv \frac1{\pi}{\rm Im}G_{l,l}^I(\w-i\delta)$ at site $l$,
where ${\hat G}^I$ is given in eq. (\ref{eqn:GIij-Iinf}).
In $N_l^I(\w)$, effect of $\delta\Sigma$ induced around the impurity
is dropped.
At (1,0) [at (1,1)],
$N_l^I(\w)$ is larger [smaller] than $N_{\rm host}(\w)$,
which is recognized as the Friedel oscillation
 \cite{Fujimoto}.
This result is changed only slightly by $GV^0$ or $GV$ methods,
which are not shown in \ref{fig:DOS}.
However, the DOS given by the $GV^I$-method is much smaller than
$N_l^I(\w)$ in under-doped systems, 
because Im$\delta\Sigma$ takes a large value around the impurity.
The Green function is given by 
\begin{eqnarray}
{\hat G}_{\rm [GVI]}&=&
 \left([{\hat G}^I]^{-1}-{\hat \delta\Sigma}_{\rm [GVI]} \right)^{-1},
\end{eqnarray}
which are easily obtained by eq.(\ref{eqn:GIab-Iinf})
in the present numerical study.
As shown in Fig. \ref{fig:DOS}, the local DOS given by the $GV^I$-method
is strongly suppressed, especially at $(1,0)$.
This suppression becomes more prominent for $U=5$.
The reason for this suppression is
the extremely short quasiparticle lifetime,
which is caused by the huge
local spin susceptibility ${\hat \chi}^{Is}$ around the impurity.
We will discuss the impurity effect on the 
spin susceptibility in the next subsection.

\begin{figure}
\begin{center}
\epsfig{file=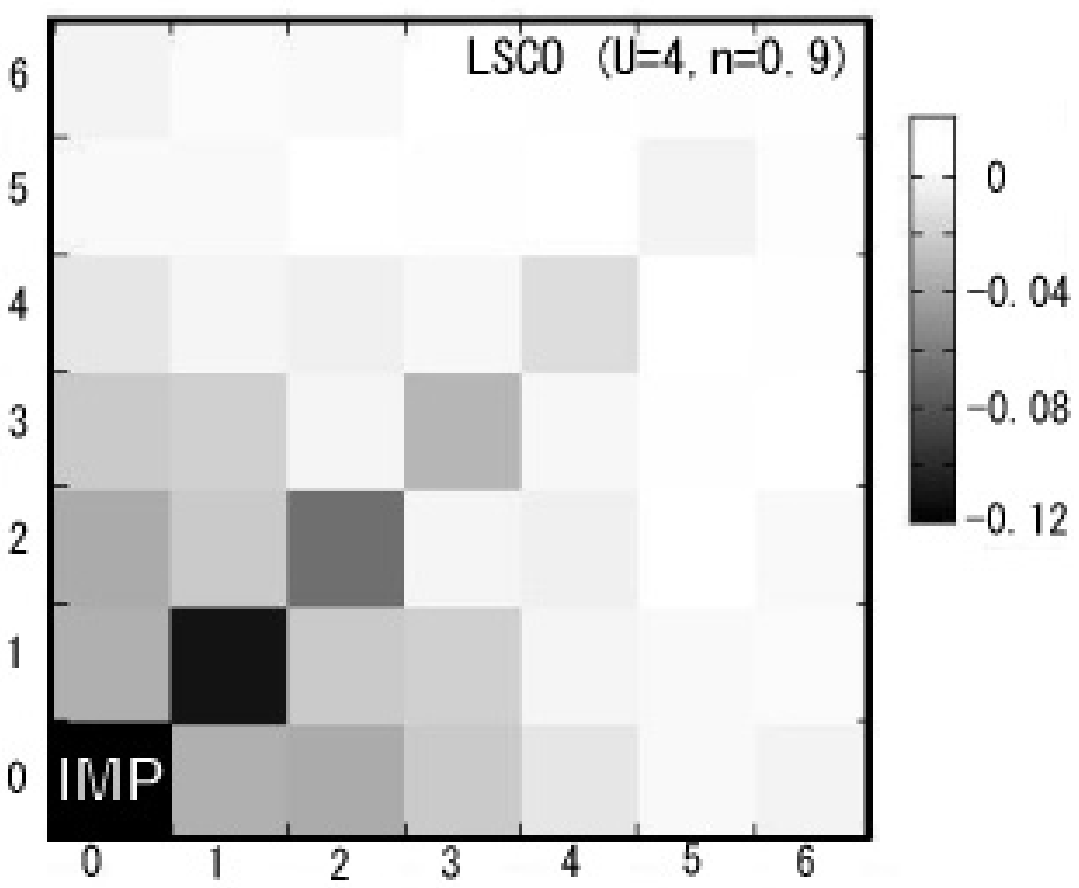,width=8cm}
\epsfig{file=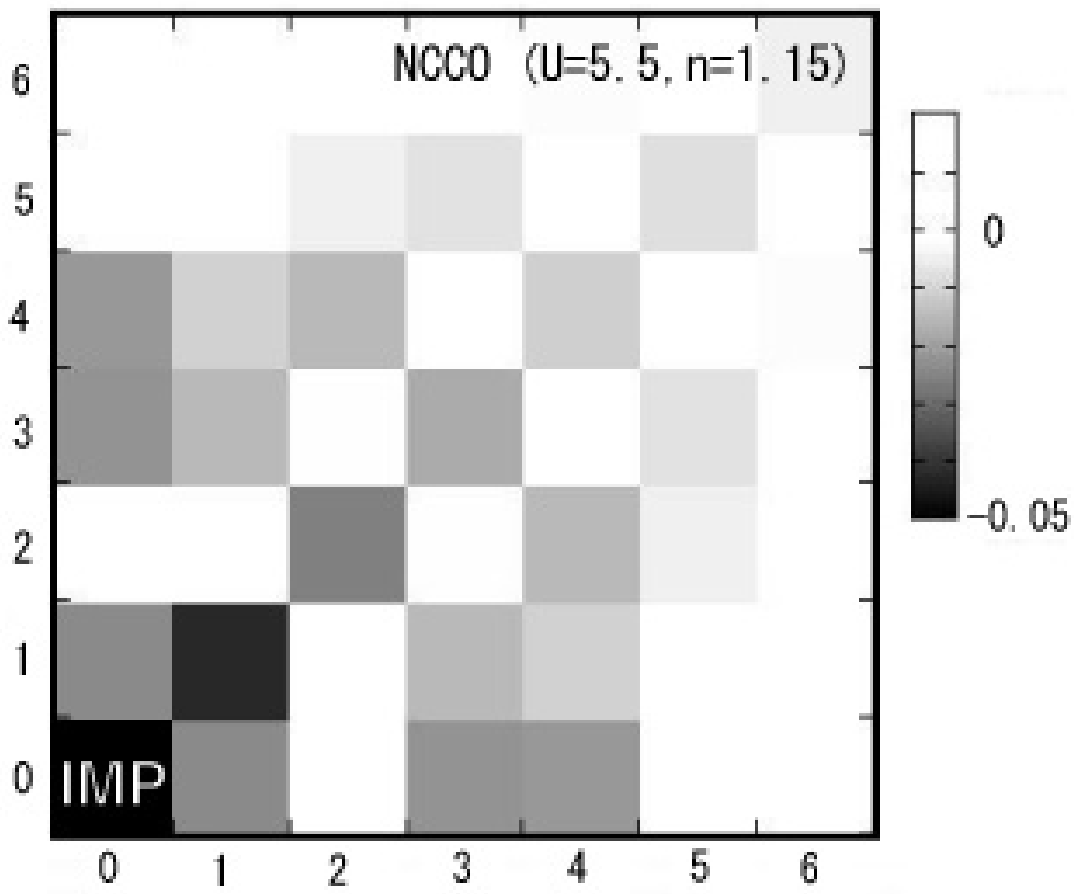,width=8cm}
\end{center}
\caption{
The reduction of the local DOS at the Fermi level around the impurity 
given by the $GV^I$-method, both for LSCO ($n=0.9$, $U=4$) and 
for NCCO ($n=1.15$, $U=5.5$) at $T=0.02$, respectively.
In the host, $N_{\rm host}(0)=0.322$ for LSCO and 
$N_{\rm host}(0)=0.217$ for NCCO, respectively.
}
\label{fig:DOS2D}
\end{figure}
Figure \ref{fig:DOS2D}
shows the reduction of the local DOS at the Fermi level
obtained by the $GV^I$-method,
both for hole-doped and electron-doped systems at $T=0.02$.
We see that the DOS is prominently suppressed in a wide region 
around the impurity, especially along the diagonal axis.
The suppression of the DOS is also recognized along
$x$, $y$-axis, which is caused by the 
enhanced quasiparticle damping, Im$\delta\Sigma(-i\delta)$,
given by the $GV^I$-method.
The strong suppression of the DOS around the impurity site
is consistent with the ``Swiss cheese structure'' 
observed in the STM measurement \cite{Pan}.
For LSCO, the radius of the Swiss cheese hole in Fig. 
\ref{fig:DOS2D} is about 3a (a being the lattice spacing).
It increases further for $U=5$, approximately 
in proportion to the AF correlation length 
$\xi_{\rm AF} (\propto \sqrt{T})$.
Because quasiparticle lifetime is extremely short in the 
Swiss cheese hole, small number of impurities will
induce the huge residual resistivity and the prominent
reduction of $T_{\rm c}$ in under-doped systems.
In \S \ref{sec:rho}, we will calculate the residual resistivity
and confirm this expectation.

\subsection{Static Spin Susceptibilities}
\label{sec:kaiS}
\begin{figure}
\begin{center}
\epsfig{file=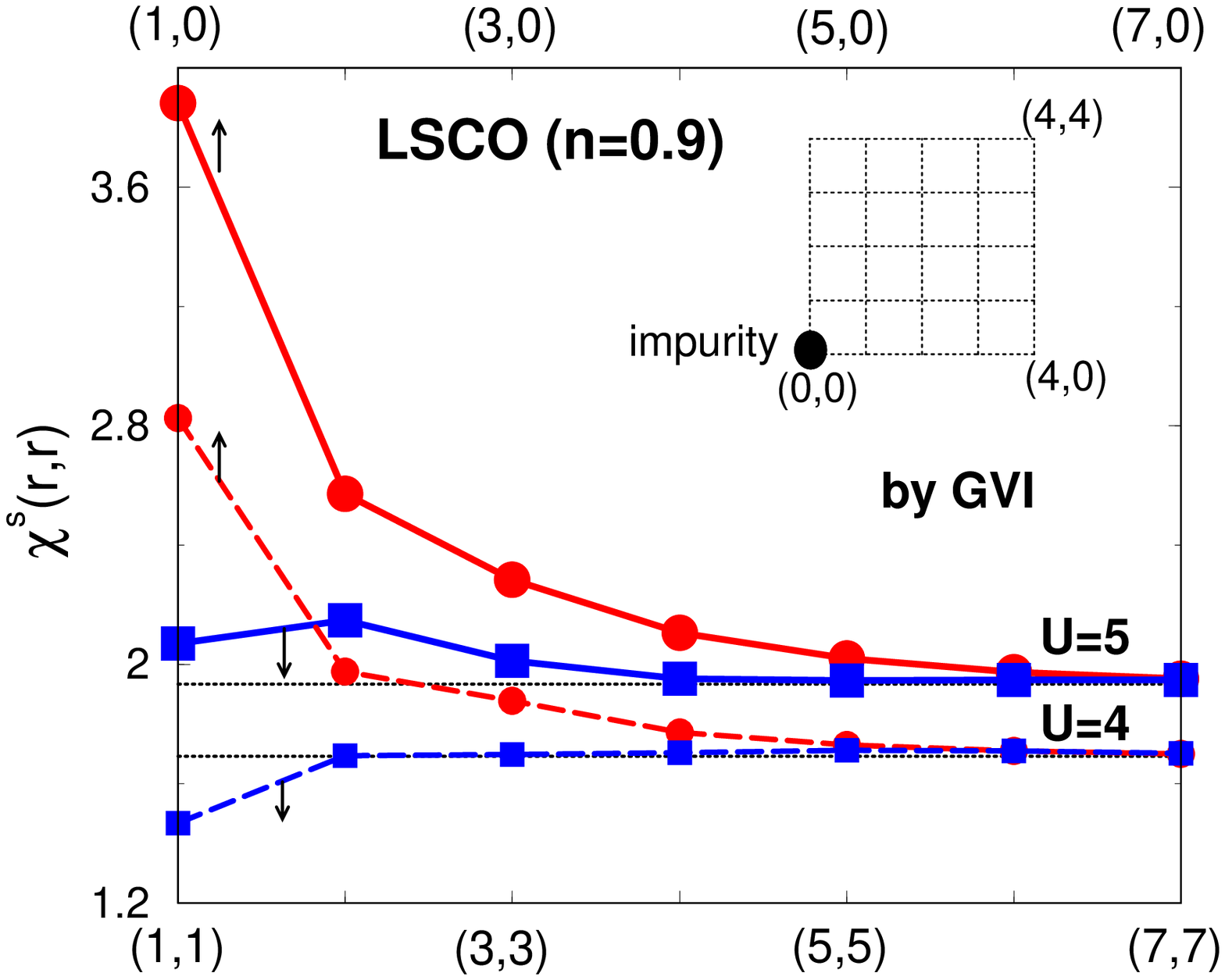,width=7cm}
\epsfig{file=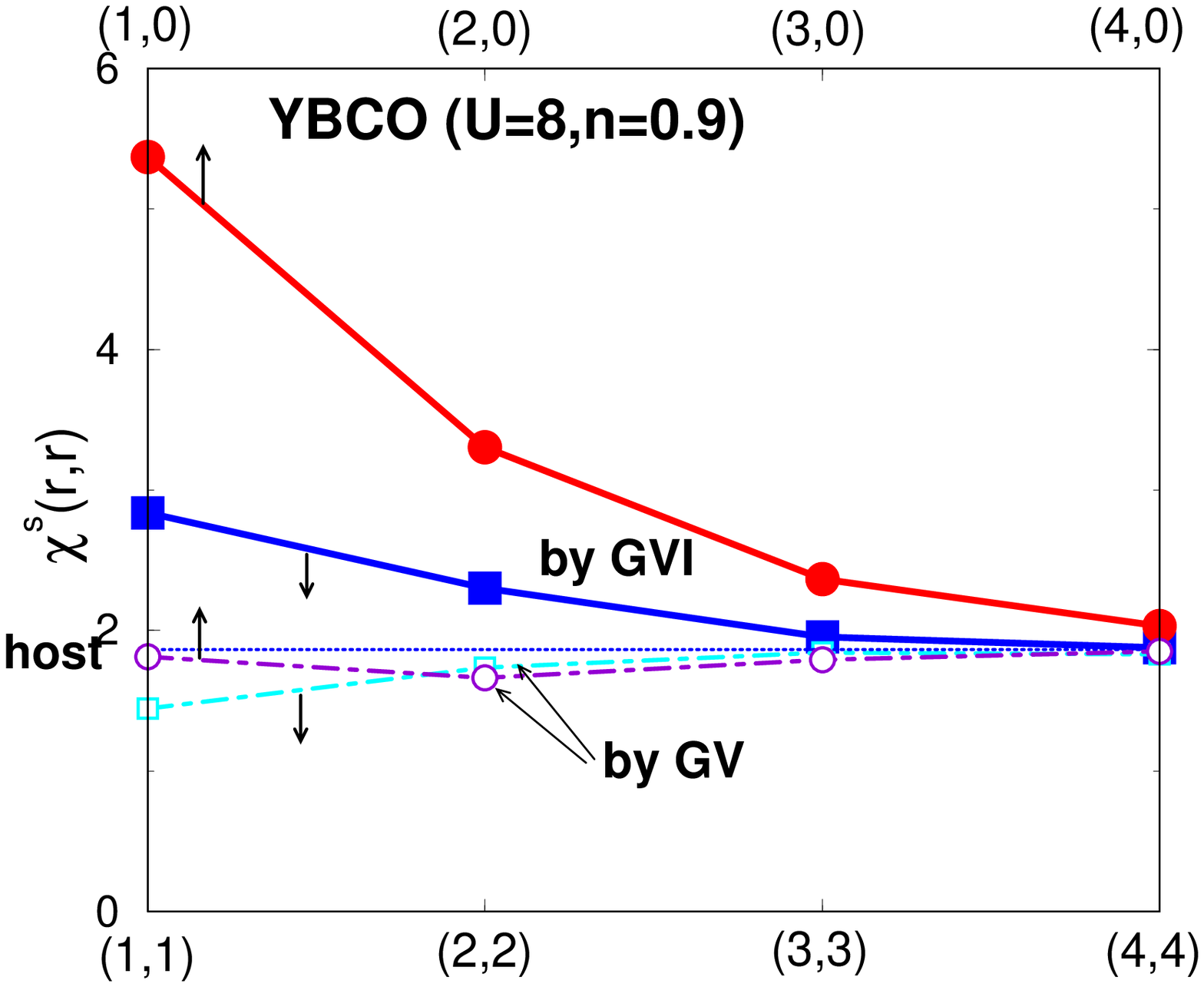,width=7cm}
\epsfig{file=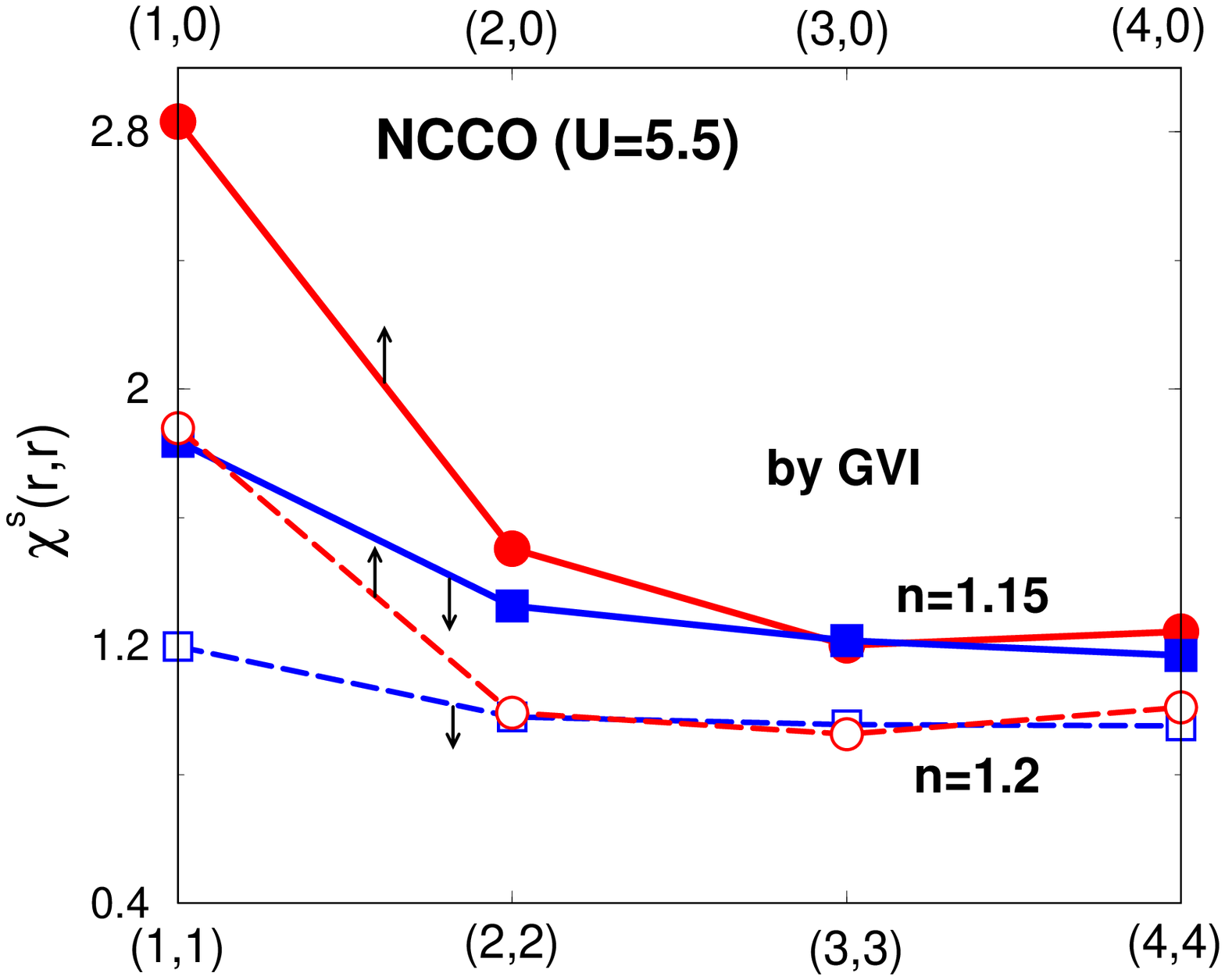,width=7cm}
\end{center}
\caption{(Color online) 
Obtained local spin susceptibility $\chi^s({\bf r},{\bf r})$
for LSCO, YBCO and NCCO at $T=0.02$.
Around the impurity site,
$\chi^s({\bf r},{\bf r})$ is strongly enhanced in the $GV^I$-method, 
whereas it slightly decreases in the $GV$-method.
The former result is consistent with experiments.
}
\label{fig:kaiS}
\end{figure}
Figure \ref{fig:kaiS} shows the obtained local spin susceptibility
$\chi^s({\bf r},{\bf r})$
along $(1,0)$ and $(1,1)$ directions at $T=0.02$.
We see that $\chi^s({\bf r},{\bf r})$ 
is significantly enhanced around the impurity
as $U$ increases, or as $n$ approaches unity.
The radius of area where $\chi^s$ is enhanced
is about $3\sim 4$a, which would corresponds to
the Swiss cheese hole in the DOS.
This result is consistent with the impurity effect
in under-doped HTSC's observed by NMR measurements.
On the other hand, an opposite result is given by the $GV$-method,
which is not reliable as discussed above.
Note that we have checked that all the eigenvalues of 
$1-U{\hat \Pi} -U{\hat {\delta \chi}}^{s*}$ in eq.(\ref{eqn:chis-I2}) 
are positive in the present numerical study.

\begin{figure}
\begin{center}
\epsfig{file=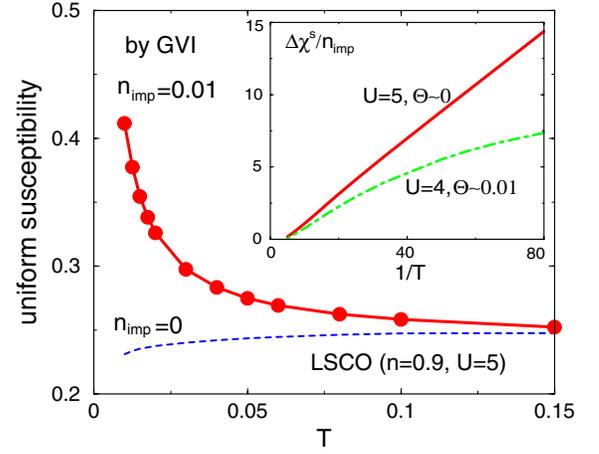,width=7.5cm}
\end{center}
\caption{(Color online) 
$T$ dependence of the uniform spin 
susceptibility given by the $GV^I$-method
in the presence or the absence of impurities.
$\Theta$ represents the Weiss temperature.
}
\label{fig:kaiS-T}
\end{figure}

Figure \ref{fig:kaiS-T} represents the 
uniform susceptibility 
$\chi_{\rm uniform}^s= N^{-2}\sum_{i,j}\chi_{i,j}^{Is}(0)$
for LSCO ($n=0.9$, $U=5$) given by the $GV^I$-method.
$n_{\rm imp}$ is the concentration of the nonmagnetic impurity.
Here, each impurity is assumed to be independent.
We see that the uniform susceptibility without impurity
decreases slightly at lower temperatures, which corresponds
to the ``weak pseudo-gap behavior'' in HTSC's
above the strong pseudo-gap temperatures, $T^\ast \sim200$K.
Surprisingly, Fig. \ref{fig:kaiS-T} shows that 
a nonmagnetic impurity induces an approximate Curie-Weiss 
like uniform susceptibility
$\Delta\chi \approx n_{\rm imp}\cdot \mu_{\rm eff}^2/3(T+\Theta)$.
For $U=5$, $\mu_{\rm eff} = 0.74\mu_{\rm B}$ and $\Theta=0$,
which means that 42\% of a magnetic moment of spin-$\frac12$
($\mu_{\rm eff} = 1.73\mu_{\rm B}$) is induced by a
single nonmagnetic impurity.
A similar behavior is obtained for YBCO in the present study.
The obtained induced moment is slightly smaller than
the experimental value $\mu_{\rm eff} \sim 1\mu_{\rm B}$
in YBa$_2$Cu$_3$O$_{6.66}$ ($T_{\rm c}\approx60K$)
 \cite{Alloul99-2}.
In the present calculation,
a simple relation $\Delta\chi \propto \chi_Q^0$
predicted by previous theoretical studies
 \cite{Bulut01,Bulut00,Ohashi,Prelovsek}
approximately holds.
The obtained $\mu_{\rm eff}$ is the present study is 
much larger and consistent with experiments.

\begin{figure}
\begin{center}
\epsfig{file=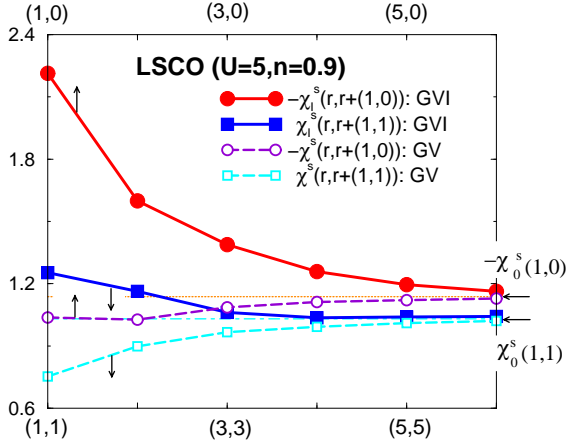,width=7.5cm}
\end{center}
\caption{(Color online) 
Nonlocal spin susceptibilities $\chi^s({\bf r},{\bf r'})$
for LSCO at $T=0.02$ given by the $GV$ and $GV^I$-methods.
Around the impurity site, the staggered susceptibility is enhanced 
in the $GV^I$-method, whereas it decreases in the $GV$-method.
The former result is consistent with experiments.
}
\label{fig:kaiS-AF}
\end{figure}

Figure \ref{fig:kaiS-AF} shows the nonlocal spin susceptibility 
around the impurity given by the $GV^I$-method,
$\chi^s({\bf r}, {\bf r}')$.
This result means that staggered susceptibility given by the 
$GV^I$-method is strongly enhanced around the impurity.
On the other hand, it is slightly suppressed in the $GV$-method.
We consider that the former result is correct whereas
the latter result is an artifact of the $GV$-method
because of the lack of the vertex corrections.
(see \S \ref{sec:VC}.)


Here, we discuss the origin of the enhancement of $\chi^{Is}$:
Reflecting the large $N_l^I(\w)$ at $l=(1,0)$,
the absolute value of $\Pi_{i,j}^I(0)$ in the $GV^I$-method
is strongly enhanced especially for $i=(1,0)$ and
(a) $j=(1,0)$, (b) $j=(1,1)$, and (c) $j=(2,1)$.
In the case of LSCO ($U=4$, $n=0.9$) at $T=0.02$,
$\Pi_{i,j}^I(0)$  [$\Pi_{i-j}^0(0)$] becomes
$0.172$  [$0.160$] for (a), $-0.160$ [$-0.100$] for (b), and 
$0.104$ [$0.0784$] for (c).
This fact will give the enhancement of 
${\hat \chi}^{Is}={\hat \Pi}^I (1-U{\hat \Pi}^I)^{-1}$
around the impurity site 
since eigenvalues of $(1-U{\hat \Pi}^I)$ become smaller.
In contrast,
an ``on-site'' nonmagnetic impurity does not give
the enhancement of spin susceptibility in the RPA
 \cite{Bulut01,Bulut00,Ohashi}.
The reason would be that the 
strong reduction of ${\hat \Pi}^0)$ due to thermal and 
quantum fluctuations are well described in the FLEX approximation.
Then, the reduction of fluctuations due to an impurity
would give rise to the enhancement of susceptibilities.

In summary, we find that
both local and staggered spin susceptibilities are 
increased around the impurity site, 
within the radius of about 3a $(\sim \xi_{\rm AF})$ at $T=0.02$.
Similar results were obtained for both the hole-doped systems 
(YBCO and LSCO) and the electron-doped ones (NCCO),
rather insensitive to model parameters.
Moreover, similar impurity effects were obtained in the 
$t$-$J$ model, by exact diagonalization studies
 \cite{Ziegler,Poilblanc}.
As a result, the obtained impurity effects in this section
will be universal in systems close to the AF-QCP.

\section{Feedback Effect and Vertex Corrections}
\label{sec:VC}
In previous sections, we show that 
the local spin susceptibility given by the $GV^I$-method,
$\chi^{Is}$, is prominently enhanced around the impurity.
However, $\chi^{Is}$ could be modified
if we go beyond the $GV^I$-method because the induced
susceptibility around the impurity, 
$\Delta\chi \equiv \chi^{Is}-\chi^{0s}$, changes the
susceptibility itself, in the form of 
the self-energy correction ($\delta\Sigma$) 
and the vertex correction (VC).
We call this self-interaction effect ``the feedback effect''.
In the feedback effect,
the susceptibility is enhanced by the VC whereas
it is reduced by $\delta\Sigma$.
In the $GV$-method, where only the latter effect
is taken into account,
the local susceptibility becomes smaller than the host's value.
This inconsistent result suggests the importance of the VC's.
In the present section,
we study the VC's for the spin susceptibility,
and show that the VC's almost cancel the self-energy correction.
We find that the total feedback effect is very small 
in the $GV$-method with VC's ($GV$+VC-method):
The obtained spin susceptibility 
is similar to that by the $GV^I$-method.
Therefore, we conclude that the $GV^I$-method is superior to the $GV$-method.

\begin{figure}
\begin{center}
\epsfig{file=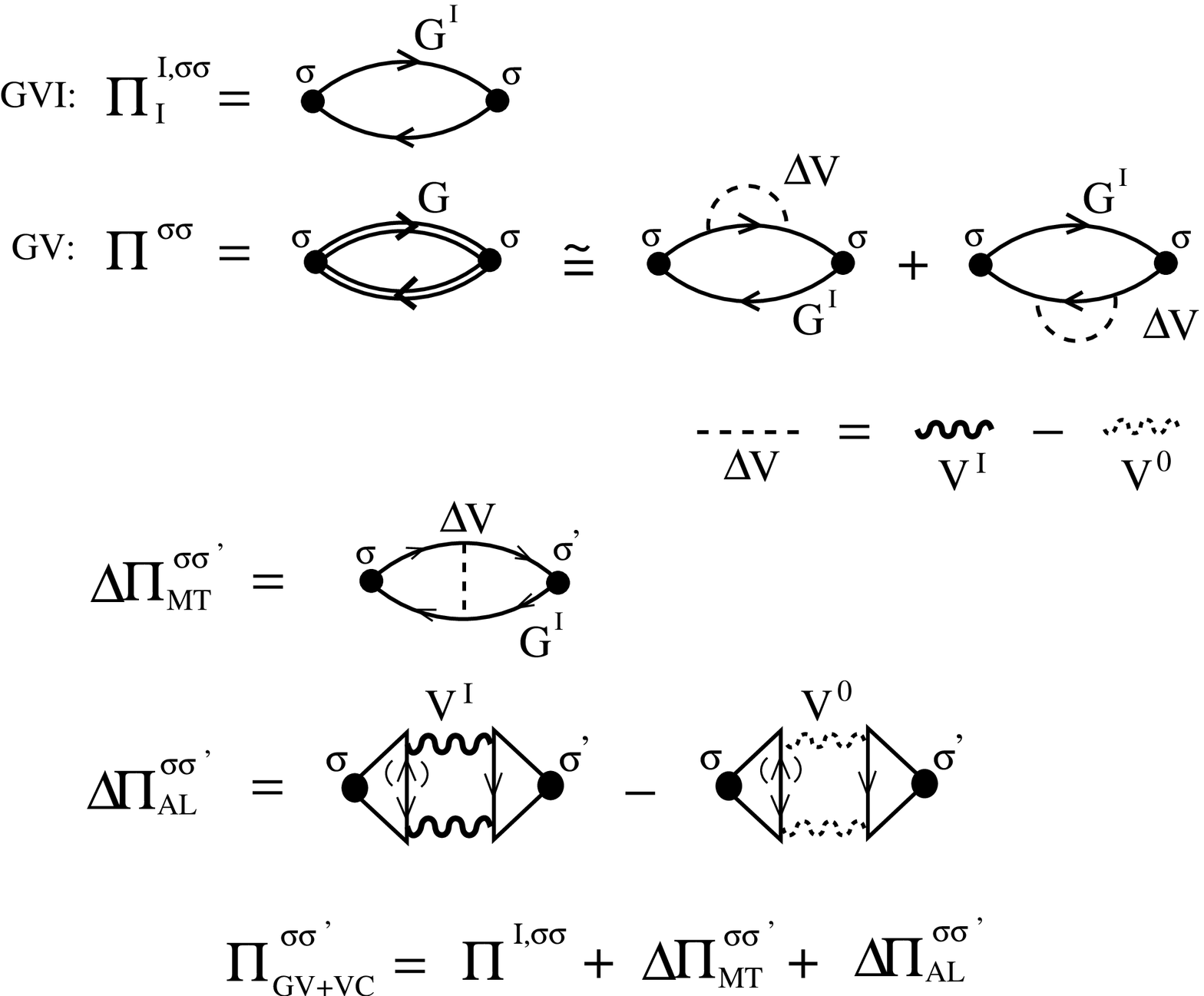,width=8cm}
\end{center}
\caption{Irreducible susceptibilities given by the 
$GV^I$, $GV$ and $GV$+VC methods, respectively.
}
\label{fig:VC}
\end{figure}

The irreducible susceptibility given by $GV^I$-method
($\Pi^{I,\s\s}$) and $GV$-method ($\Pi^{\s\s}$) are expressed 
in eqs.(\ref{eqn:Pi-GV}) and (\ref{eqn:Pi-GVI}), respectively.
Hereafter, we discuss the feedback effect for the 
irreducible susceptibility perturbatively
with respect to $\Delta {\hat V} = {\hat V}^I-{\hat V}^0$.
In this respect, $\Pi^{\s\s}$ can be expanded with respect to 
$\Delta {\hat V}$ as
\begin{widetext}
\begin{eqnarray}
\Pi^{\s\s}_{i,j}(0) &\approx& \Pi^{I,\s\s}_{i,j}
 -2T\sum_{i',j',\e_m,\e_n} G_{j',j}^I(\e_n)G_{j,i}^I(\e_n)G_{i,i'}^I(\e_n)
  \cdot \Delta V_{i',j'}(\e_n-\e_m)G_{i',j'}^I(\e_m) ,
 \label{eqn:Pi-exp}
\end{eqnarray}
up to the lowest order, which is expressed in Fig. \ref{fig:VC}.
We have checked numerically that eq. (\ref{eqn:Pi-exp})
is satisfied well.

Next, we study the VC for irreducible susceptibility
up to the second-order with respect to $\Delta {\hat V}$.
The lowest order term, which we call the Maki-Thompson (MT) term
customarily, is given by
\begin{eqnarray}
\Delta \Pi_{\rm MT}^{\uparrow\uparrow}(i,j)
&=& -\frac{U^2T^2}{2} \sum_{i',j';\e_m,\e_n}F(i,j,i',j';\e_m,\e_n)
 \left(\Delta\chi_{i',j'}^s(\e_m-\e_n) 
 +\Delta\chi_{i',j'}^c(\e_m-\e_n) \right)
 \label{eqn:Pi-MT},
 \\
\Delta \Pi_{\rm MT}^{\uparrow\downarrow}(i,j)
&=& -\frac{U^2T^2}{2} \sum_{i',j';\e_m,\e_n}F(i,j,i',j';\e_m,\e_n)
 \Delta\chi_{i',j'}^s(\e_m-\e_n)
 \label{eqn:Pi-MT2},
 \\
\Delta\chi_{i,j}^{s,c} &=& \chi_{i,j}^{Is,c}-\chi_{i,j}^{0s,c} ,
 \\
F(i,j,i',j';\e_m,\e_n) &=&
 G_{j',i}^I(\e_m)G_{i,i'}^I(\e_m)G_{i',j}^I(\e_n)G_{j,j'}^I(\e_n) ,
\end{eqnarray}
which is shown in Fig. \ref{fig:VC}.
We also discuss the second-order term given as
\begin{eqnarray}
\Delta \Pi_{\rm AL}^{\s\s'}(i,j)
&=& T \sum_{i_1,i_2,j_1,j_2;\w_l}
 F'(i,i_1,i_2;\w_l)(F'(j,j_1,j_2;\w_l)+F'(j,j_1,j_2;-\w_l))
 \nonumber \\
& &\times (X_{\s\s'}^I(i_1,i_2,j_1,j_2;\w_l) - X_{\s\s'}^0(i_1,i_2,j_1,j_2;\w_l))
 \label{eqn:Pi-AL},
 \\
F'(i,i_1,i_2;\w_l) &=& T\sum_{\e_n}
 G_{i_2,i}^I(\e_n)G_{i,i_1}^I(\e_n)G_{i_1,i_2}^I(\e_n+\w_l) ,
 \\
X_{\uparrow\uparrow}^\xi(i_1,i_2,j_1,j_2;\w_l)
&=& \frac{5U^4}{4}\chi_{i_1,j_1}^{\xi s}\chi_{i_2,j_2}^{\xi s}
 + \frac{U^3}{4} \chi_{i_1,j_1}^{\xi s}(U\chi_{i_2,j_2}^{\xi c}+4\delta_{i_2,j_2})
 +\frac{U^3}{4}\chi_{i_2,j_2}^{\xi s}(U\chi_{i_1,j_1}^{\xi c}+4\delta_{i_1,j_1})
 \nonumber \\
& & + \frac{U^4}{4} \chi_{i_1,j_1}^{\xi c}\chi_{i_2,j_2}^{\xi c} ,
 \\
X_{\uparrow\downarrow}^\xi(i_1,i_2,j_1,j_2;\w_l)
&=& \frac{U^4}{4}(\chi_{i_1,j_1}^{\xi s}-\chi_{i_1,j_1}^{\xi c})
 (\chi_{i_2,j_2}^{\xi s}-\chi_{i_2,j_2}^{\xi c})
+ \frac{U^3}{2}\delta_{i_1,j_1} (\chi_{i_2,j_2}^{\xi s}-\chi_{i_2,j_2}^{\xi c})
 \nonumber \\
& &+ \frac{U^3}{2}\delta_{i_2,j_2} (\chi_{i_1,j_1}^{\xi s}-\chi_{i_1,j_1}^{\xi c}) ,
\end{eqnarray}
where $\xi=0$ or $I$.
We call eq.(\ref{eqn:Pi-AL}) the Aslamazov-Larkin term.
\end{widetext}
In the FLEX approximation,
the irreducible VC's given by the Ward identity,
$\Gamma^{\rm irr}=\delta\Sigma/\delta G$,
are composed of the MT-term and the AL-term
 \cite{Bickers}.

As a result, the irreducible susceptibility
given by the ``$GV$-method with VC's up to the
second-order with respect to $\Delta\chi^{s,c}$''
($GV$+VC method) is given by
\begin{eqnarray}
\Pi^{\s\s'}(i,j)
&=&  \Pi(i,j)\delta_{\s,\s'}
\nonumber \\
& &+ \Delta \Pi_{\rm MT}^{\s\s'}(i,j)+\Delta \Pi_{\rm AL}^{\s\s'}(i,j) .
\end{eqnarray}
The spin susceptibility in the $GV$+VC method is obtained as
\begin{eqnarray}
{\hat \chi}_{GV+{\rm VC}}^s &=& {\hat \chi}_{\uparrow\uparrow}+
 {\hat \chi}_{\uparrow\downarrow} ,
 \label{eqn:chi-GVVC}
 \\
{\hat \chi}_{\uparrow\uparrow} &=&
 {\hat \Pi}^{\uparrow\uparrow}
+{\hat \Pi}^{\uparrow\uparrow}U{\hat \chi}_{\uparrow\downarrow}
-{\hat \Pi}^{\uparrow\downarrow}U{\hat \chi}_{\uparrow\uparrow},
 \label{eqn:chi-uu}
 \\
{\hat \chi}_{\uparrow\downarrow} &=&
 -{\hat \Pi}^{\uparrow\uparrow}
+{\hat \Pi}^{\uparrow\uparrow}U{\hat \chi}_{\uparrow\uparrow}
-{\hat \Pi}^{\uparrow\downarrow}U{\hat \chi}_{\uparrow\downarrow},
 \label{eqn:chi-ud}
\end{eqnarray}
%

\begin{figure}
\begin{center}
\epsfig{file=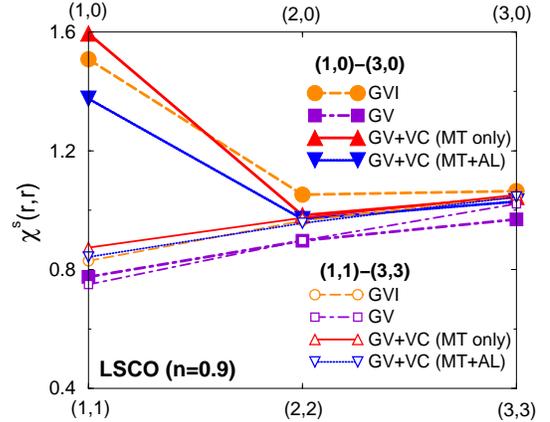,width=7cm}
\end{center}
\caption{(Color online) 
Local spin susceptibilities for $U=4$ at $T=0.02$
given by the $GV$, $GV^I$ and $GV$+VC methods, respectively.
$\chi^s({\bf r},{\bf r})$ at ${\bf r}=(1,0)$ takes an enhanced
value for both $GV^I$ and $GV$+VC methods, whereas
it decreases in the $GV$-method.
We conclude that $GV^I$-method is as good as $GV$+VC-method.
}
\label{fig:kaiS-VC}
\end{figure}

In the present numerical study, 
we calculate $\Delta \Pi_{\rm MT,AL}(\a,\b)$
only for region A in Fig. \ref{fig:region}.
Here we put $M=3$ in the present section.
Unfortunately, it is not easy to calculate all the elements of
eqs. (\ref{eqn:Pi-MT}), (\ref{eqn:Pi-MT2}) and (\ref{eqn:Pi-AL}) 
in the region A because of the huge computation time.
Therefore, we calculate $\Delta \Pi_{\rm MT,AL}(\a,\b)$
only for $|\a-\b|\le4$ in region A, and derive the static spin 
susceptibility ${\hat \chi}_{GV+{\rm VC}}^s$ by solving 
eqs.(\ref{eqn:chi-GVVC}) - (\ref{eqn:chi-ud}) for region A+B.

Figure \ref{fig:kaiS-VC} show the 
local spin susceptibility given by $GV^I$-method [${\hat \chi}^{Is}(i,i)$],
$GV$-method [${\hat \chi}^{s}(i,i)$], and $GV$+VC-method 
[${\hat \chi}_{GV+{\rm VC}}^{s}(i,i)$] for LSCO ($U=4$, $n=0.9$), 
respectively.
At $i=(1,0)$, we see that ${\hat \chi}^{Is}(i,i)$ increases whereas 
${\hat \chi}^{s}(i,i)$ decreases,
as we have shown in Fig. \ref{fig:kaiS}.
Note that we have put 
$\Pi(i,j)= \Pi_I(i,j)= \Pi_{GV+{\rm VC}}(i,j)=0$
for $|i-j|>4$ in deriving Fig. \ref{fig:kaiS-VC}
to make comparison between different methods.
Because of this fact, both 
${\hat \chi}^{Is}(i,i)$ and ${\hat \chi}^{s}(i,i)$ in Fig. \ref{fig:kaiS-VC}
are smaller than those in Fig. \ref{fig:kaiS} for $U=4$.

As shown in Fig. \ref{fig:kaiS-VC},
${\hat \chi}_{GV+{\rm VC}}^{s}(i,i)$ at $i=(1,0)$
is strongly enhanced due to the VC's,
to be comparable with ${\hat \chi}^{Is}(i,i)$.
This enhancement is brought mainly by the MT-term,
whereas the AL-term slightly reduces the spin susceptibility.
Therefore, the suppression of ${\hat \chi}^{s}$ 
in the $GV$-method, which is caused by the non-local self-energy 
correction $\delta\Sigma$, is almost recovered by the VC's.
Therefore, the $GV^I$-method gives a reliable 
spin susceptibility around the impurity, because the total
feedback effects almost cancel.

We comment that the superiority of the $GV^I$-method
over the $GV$-method and the importance of the VC's
would remind us of the $GW$ approximation
 \cite{Holm1,Holm2,Godby}.
It is a first principle calculation for the self-energy,
which is given by the convolution of the Green function $G$ and the 
screened interaction $W$ within the RPA.
In the ``fully self-consistent $GW$'',
both $G$ and $W$ are obtained self-consistently.
In the $GW_0$ scheme, on the other hand, the self-energy is 
given by $G$ and $W_0$, where $W_0$ is the screened interaction
without self-energy correction.
Reference \cite{Holm1,Holm2,Godby}
shows that the descriptions of the bandwidth reduction
and the satellite structure in the quasiparticle spectrum
are satisfactory in the $GW_0$,
whereas results given by $GW$ are much worse.
This result clearly indicates the necessity of VC's in the $GW$.

\section{Transport Phenomena}
\label{sec:transport}

In previous sections, we showed that the magnetic susceptibility 
is strongly enhanced around the impurity.
This fact will cause the strong nonlocal change in the 
self-energy around the impurity, $\delta\Sigma$.
In the present section, we will show that $\delta\Sigma$ gives rise 
to a huge residual resistivity at finite temperatures in nearly AF 
Fermi liquids, which is the most important finding in the present paper.
Moreover, a small number of nonmagnetic impurities can cause 
a ``Kondo-like'' insulating behavior ($d\rho/dT<0$) at low temperatures, 
when the system is very close to AF-QCP.
Different from a conventional single-channel Kondo effect,
the residual resistivity can be much larger than
the value for s-wave unitary scattering.
These findings naturally explain various long-standing
problems on the transport phenomena in HTSC's,
heavy fermion systems and organic superconductors.

\subsection{$T$-matrix}
\label{sec:Tmatrix}

\begin{figure}
\begin{center}
\epsfig{file=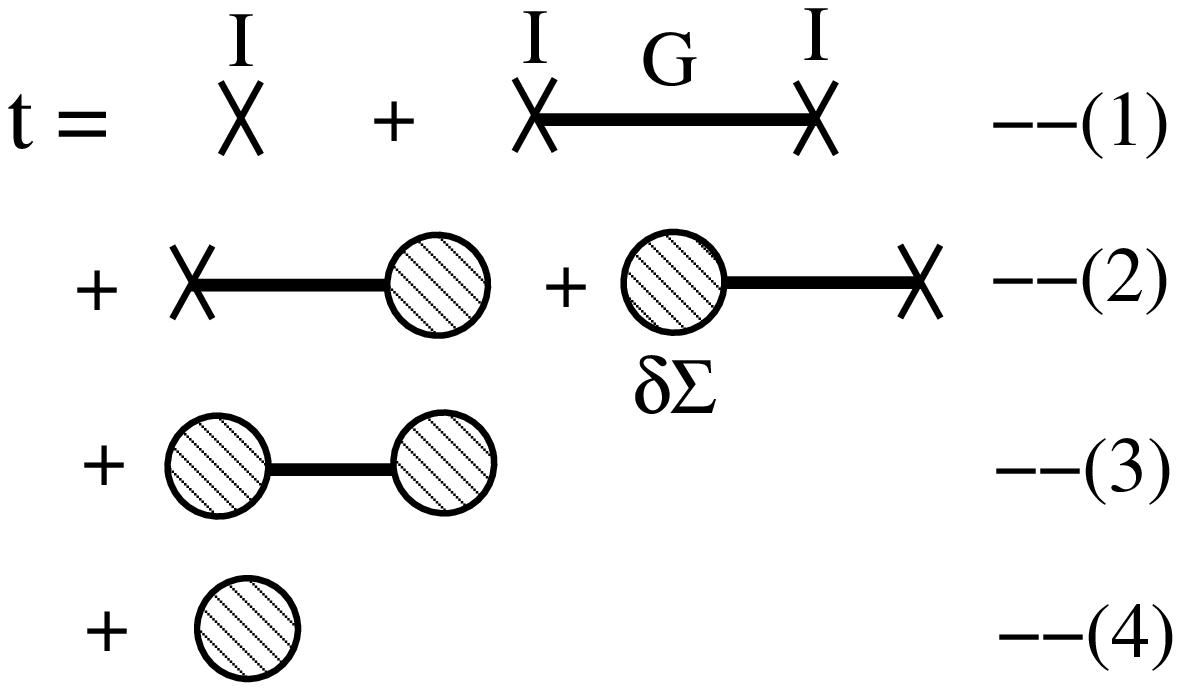,width=6cm}
\end{center}
\caption{
Diagrammatic expression for the t-matrix given 
in eqs. (\ref{eqn:t1})-(\ref{eqn:t4}).
}
  \label{fig:Tmatrix}
\end{figure}

First, we derive the expression for the $t$-matrix,
${\hat t}(\e)$, which is defined as
${\hat G}= {\hat G}^0 + {\hat G}^0 {\hat t} {\hat G}^0$.
Therefore, the $t$-matrix due to the impurity potential $I$
at center ${\bf r}=(0,0)$ is given by 
\begin{eqnarray}
t_{\a,\b}&=& \sum_{n=1}^4 t_{\a,\b}^{(n)},
 \label{eqn:t1234} \\
t_{\a,\b}^{(1)}&=&
 (I+I^2 G_{0,0}) \delta_{\a,0}\delta_{\b,0},
 \label{eqn:t1} \\
t_{\a,\b}^{(2)}&=&
 I\sum_{\nu}^{\rm A} \left( G_{0,\nu}\delta\Sigma_{\nu,\b}\cdot \delta_{\a,0}
 + \delta\Sigma_{\a,\nu}G_{\nu,0}\cdot \delta_{\b,0} \right),
 \label{eqn:t2} \\
t_{\a,\b}^{(3)}&=&
 \sum_{\nu,\mu}^{\rm A} \delta\Sigma_{\a,\nu}G_{\nu,\mu}
 \delta\Sigma_{\mu,\b},
 \label{eqn:t3} \\
t_{\a,\b}^{(4)}&=& \delta\Sigma_{\a,\b} ,
 \label{eqn:t4} 
\end{eqnarray}
where $t_{\a,\b}^{(1)} \sim t_{\a,\b}^{(4)}$ 
is schematically expressed in Fig. \ref{fig:Tmatrix}.
$\a$ and $\b$ represent sites in region A.

Here, we derive the expression for the $t$-matrix 
in the limit $I\rightarrow \infty$.
First, the second term of eq. (\ref{eqn:t1}) is rewritten as
\begin{eqnarray}
I^2 G_{0,0}&=& I^2 G_{0,0}^I + I^2\sum_{\a,\b}^{\rm A}
 G_{0,\a}^I \delta\Sigma_{\a,\b} G_{\b,0}^I
 \nonumber \\
& &+ I^2 \sum_{\a,\b,\gamma,\delta}^{\rm A}
 G_{0,\a}^I \delta\Sigma_{\a,\b} G_{\b,\gamma}
 \delta\Sigma_{\gamma,\delta}^I G_{\delta,0}^I .
\end{eqnarray}
According to eq. (\ref{eqn:GI0}),
\begin{eqnarray}
I^2 G_{0,0}^I &=& -I -\frac1{G_{0,0}^0} + O(I^{-1}),
 \\
I G_{0,\a}^I &=& -\frac{G_{0,\a}^0}{G_{0,0}^0} + O(I^{-1}) .
\end{eqnarray}
Thus, eq. (\ref{eqn:t1}) in the limit $I\rightarrow \infty$ is
given by
\begin{eqnarray}
I+I^2 G_{0,0}&=& -\frac1{G_{0,0}^0}
+ \sum_{\a\b}^{\rm A} \frac{G_{0,\a}^0 G_{\b,0}^0}{(G_{0,0})^2}
 \nonumber \\
& &\times\left( \delta\Sigma_{\a,\b} + \sum_{\gamma\delta}^{\rm A}
 \delta\Sigma_{\a,\gamma}G_{\gamma,\delta}
 \delta\Sigma_{\delta,\b} \right) .
 \label{eqn:t1-a}
\end{eqnarray}

In the same way, eq. (\ref{eqn:t2}) 
in the limit $I\rightarrow \infty$ is given by
\begin{eqnarray}
& &-\sum_\delta^{\rm A} \frac{\delta\Sigma_{\a,\delta} }{G_{0,0}^0}
 \left( G_{\delta,0}^0
 +\sum_{\xi\eta}^{\rm A} G_{\delta,\xi}\delta\Sigma_{\xi,\eta}
 G_{\eta,0}^0 \right) \delta_{0,\b} 
 \nonumber \\
& &\ \ \ \ 
 + \langle \a \leftrightarrow \b \rangle .
\end{eqnarray}
Note that both eqs. (\ref{eqn:t3}) and (\ref{eqn:t4})
contain $I$ only through $\delta\Sigma$,
which does not diverge even when $I=\infty$.

To study the effect of the impurity on the transport phenomena, 
we take the average of the t-matrix 
with respect to the position of the impurity.
The obtained result is
\begin{eqnarray}
T_l &\equiv& \sum_\a^{\rm A} t_{l+\a,\a}
 \nonumber \\
&=& -\frac1{G_{0,0}^0}\left( 
 1- \frac1{G_{0,0}^0}\sum_{\a\b}^{\rm A} D_{0,\a}^{0} 
 (\delta_{\a,\b}+D_{\a,\b}) G_{\b,0}^0 \right) \delta_{l,0}
 \nonumber \\ 
 \label{eqn:T1} \\
& &- \frac2{G_{0,0}^0}\sum_{\a}^{\rm A}
 D_{0,\a}^{0} \left( \delta_{\a,l} + D_{\a,l} \right)
 \label{eqn:T2} \\
& &+ \sum_{\a\b\gamma}^{\rm A} \delta\Sigma_{l+\a,\b}G_{\b,\gamma}
 \delta\Sigma_{\gamma,\a}
 \label{eqn:T3} \\
& &+ \sum_{\a}^{\rm A} \delta\Sigma_{l+\a,\a} ,
 \label{eqn:T4}
\end{eqnarray}
where eqs. (\ref{eqn:T1}), (\ref{eqn:T2}),(\ref{eqn:T3})
and (\ref{eqn:T4}) come from eqs. (\ref{eqn:t1}), (\ref{eqn:t2}),
(\ref{eqn:t3}) and (\ref{eqn:t4}), respectively.
$D_{\a,\b}^{0}$ is given in eq. (\ref{eqn:D0}).
Similarly, $D_{\a,\b}$ is defined as 
\begin{eqnarray}
D_{\a,\b}&=& \sum_\gamma^{\rm A} G_{\a,\gamma}\delta\Sigma_{\gamma,\b}
 = \sum_\gamma^{\rm A} \delta\Sigma_{\b,\gamma} G_{\gamma,\a} .
 \label{eqn:D}
\end{eqnarray}
Note that
$D_{\a,0}=D_{0,\a}=0$ and $D_{\a,0}^{0}=0$, whereas 
$D_{0,\a}^{0}\ne 0$.
After the analytic continuation of $T_l(\e_n)$,
the quasiparticle damping rate 
(without the renormalization factor) due to the impurity
is given by 
 \cite{Eliashberg,Langer}
\begin{eqnarray}
\gamma_\k^{\rm imp}(\e) &=& \frac{n_{\rm imp}}{N^2} 
 \sum_l {\rm Im}T_l(\e-i\delta)e^{i\k \cdot {\bf r}_l} ,
\label{eqn:gammak}
\end{eqnarray}
where $n_{\rm imp}$ is the density of impurities.

\begin{figure}
\begin{center}
\epsfig{file=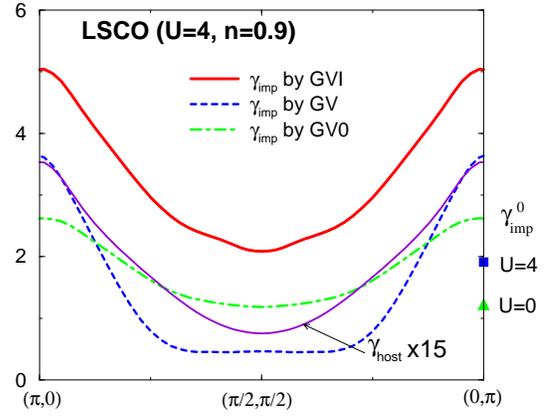,width=7cm}
\end{center}
\caption{(Color online) 
$\k$-dependence of $\gamma_\k^{\rm imp}(0)$
per $n_{\rm imp}$ at $T=0.02$, along the Fermi line.
The shape of the Fermi line is shown in ref.\cite{Kontani-Hall}.
$\gamma_{\rm host}={\rm Im}\Sigma_{\k}^0(-i\delta)$
for the host system.
}
\label{fig:Gamma-comp}
\end{figure}

Figure \ref{fig:Gamma-comp} shows $\gamma_\k^{\rm imp}(0)$
along the Fermi surface for LSCO at $T=0.02$, 
obtained by $GV^0$, $GV$ and $GV^I$-methods.
Here, we put $n_{\rm imp}=1$.
$\gamma_\k^{\rm host}={\rm Im}\Sigma_\k^0(-i\delta)$
in the host system given by the FLEX approximation.
Filled square and triangular represents
\begin{eqnarray}
\gamma_{\rm imp}^0(\e)\equiv 
 -{\rm Im}\frac{n_{\rm imp}}{G_0^0(\e-i\delta)} ,
 \label{eqn:gamma0}
\end{eqnarray}
for $n_{\rm imp}=1$,
where $G_0^0(\e)=\frac1{N^2}\sum_\k G_\k^0(\e)$
is the local Green function of the host.
Equation (\ref{eqn:gamma0}) is a well-known expression
for the quasiparticle damping rate due to s-wave unitary impurities.
In fact, $\gamma_{\rm imp}^0$ is derived from eq. (\ref{eqn:gammak})
by putting $\delta\Sigma_{\a,\b}=-\Sigma_{\a-\b}^0 
\cdot\delta_{\a\b,0}$ in eqs.(\ref{eqn:T1})-(\ref{eqn:T4}).
When the particle-hole symmetry is approximately satisfied 
around the Fermi level, eq.(\ref{eqn:gamma0}) becomes
$\gamma_{\rm imp}^0(\e)= n_{\rm imp}/\pi N_{\rm host}(\e)$,
where $N_{\rm host}(\e)={\rm Im}G_0^0(\e-i\delta)/\pi$ 
is the DOS of the host system.

In each method ($GV$, $GV^I$ and $GV^0$), $\gamma_\k^{\rm imp}$ 
has a strong $\k$-dependence similar to $\gamma_\k^{\rm host}$, 
as shown in Fig. \ref{fig:Gamma-comp}.
As a result, the structure of the ``hot spot'' and the
``cold spot'', which are located around $(\pi/2,\pi/2)$
and $(\pi/,0)$ respectively, is not smeared out
by strong non-magnetic impurities.
This highly nontrivial result is brought by the 
$\k$-dependence of $\delta\Sigma_\k$.
This finding strongly suggests that the
enhancement of the Hall coefficient near the AF-QCP,
which is brought by the strong back-flow 
(current vertex correction) around the cold spot 
 \cite{Kontani-Hall,Kontani-rev}, 
does not decrease due to the strong impurities.
In fact, the Hall coefficient for under-doped YBCO
slightly increases by the doping of non-magnetic impurities
 \cite{Ong}.

On the other hand, 
in the case of weak impurities where
the Born approximation is reliable,
$\gamma_\k^{\rm imp}$ should be almost $\k$-independent.
Thus, the structure of the ``hot/cold spots'' will be
smeared out by (large numbers of) weak non-magnetic impurities.
By this reason, the enlarged Hall coefficient near the AF-QCP,
which is brought by the back-flow around the cold spot,
is reduced by weak non-magnetic impurities
 \cite{future}.
This theoretical result would be able to explain 
the reduction of $R_{\rm H}$ in CeCoIn$_5$ at very low temperatures
 \cite{Ce115,Ce115-2}.

\begin{figure}
\begin{center}
\epsfig{file=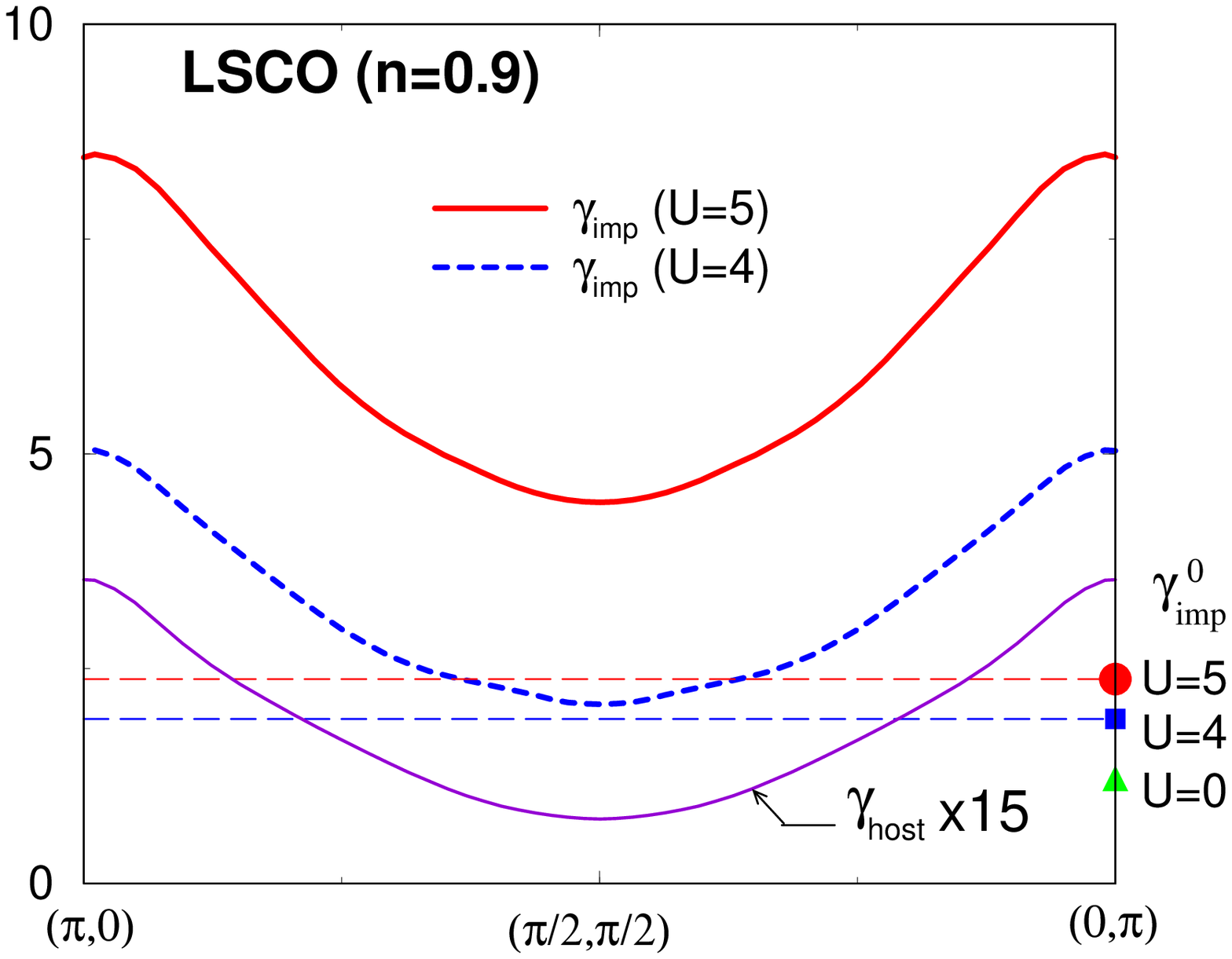,width=6.5cm}
\epsfig{file=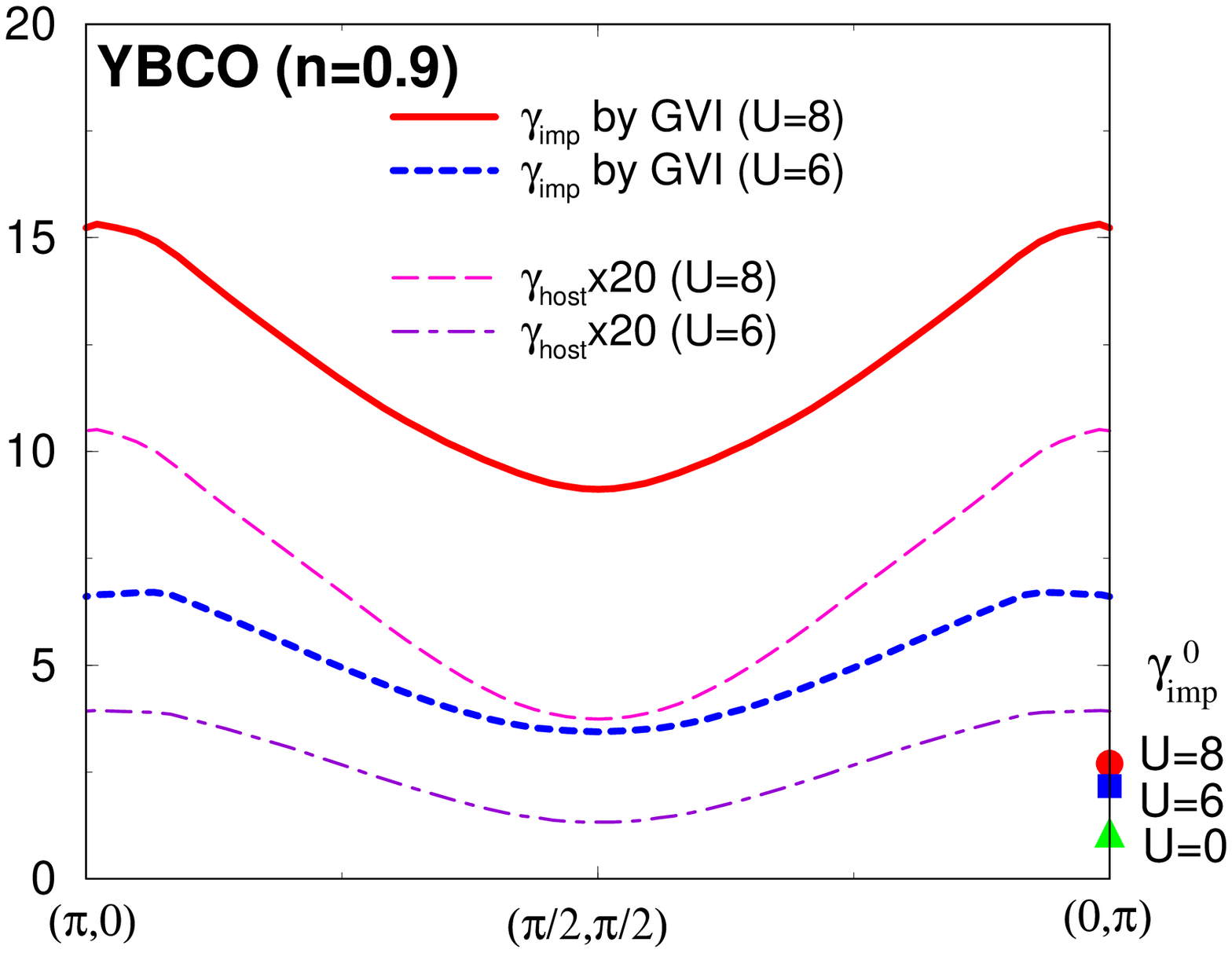,width=6.5cm}
\epsfig{file=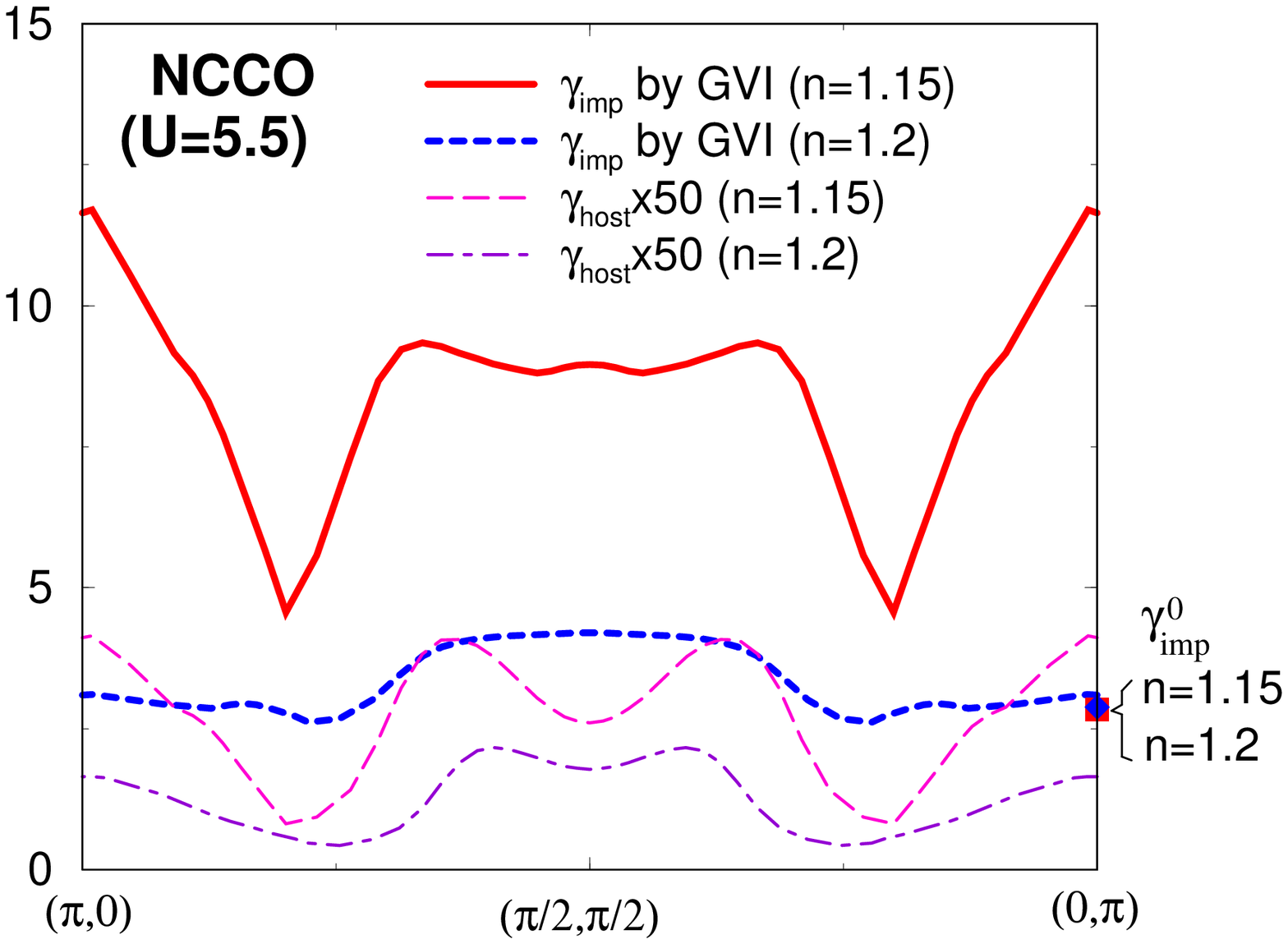,width=6.5cm}
\end{center}
\caption{(Color online) 
$\gamma_\k^{\rm imp}(0)$ 
per $n_{\rm imp}$ at $T=0.02$, along the Fermi line.
They are given by the $GV^I$-method.
Both $\gamma_\k^{\rm imp}$ and $\gamma_{\rm host}$ show
similar $\k$-dependence, which means that the
hot/cold spot structure is maintained against the doping of 
strong nonmagnetic impurities.
Moreover, $\gamma_\k^{\rm imp}$ is much larger than 
$\gamma_{\rm imp}^0$ when AF fluctuations are strong.
}
\label{fig:Gamma}
\end{figure}

Another important finding
is that $\gamma_\k^{\rm imp}$ given by the $GV^I$-method 
becomes larger than $\gamma_{\rm imp}^0$, due to the
non-local scattering (non s-wave scattering) given by $\delta\Sigma$.
Figure \ref{fig:Gamma}
shows $\gamma_\k^{\rm imp}(0)$ for $n_{\rm imp}=1$
given by the $GV^I$-method, at $T=0.02$ for LSCO, YBCO and NCCO.
In both hole and electron-doped systems, the hot/cold spot structure 
in the host system remains even in the presence of impurities.
The absolute value of $\gamma_\k^{\rm imp}(0)$
increases drastically when the system is close to the AF-QCP.
as shown in \ref{fig:Gamma}.
This finding gives the explanation for the huge residual resistivity 
in metals near the AF-QCP, which has been
a long-standing problem in strongly correlated electron systems.
In contrast, $\gamma_\k^{\rm imp}$ given by the $GV^0$-method
is comparable to $\gamma_{\rm imp}^0$, because the enhancement of
${\hat \chi}^s$ around the impurity is not taken into account.
Also, $\gamma_\k^{\rm imp}$ by the $GV$-method is much 
smaller than $\gamma_{\rm imp}^0$.
This result will be an artifact of the $GV$-method,
as discussed in \S \ref{sec:VC}.

\begin{figure}
\begin{center}
\epsfig{file=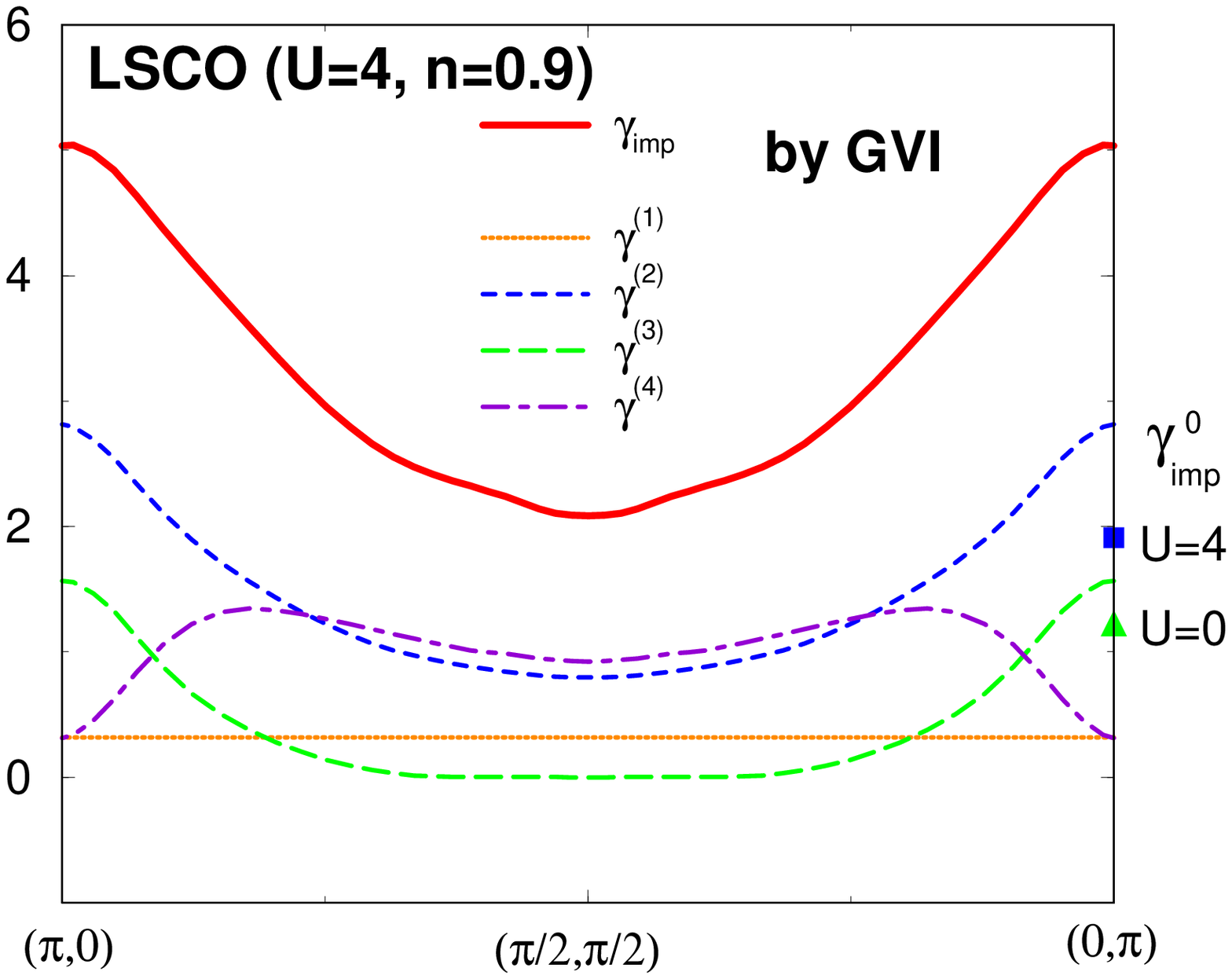,width=6.5cm}
\epsfig{file=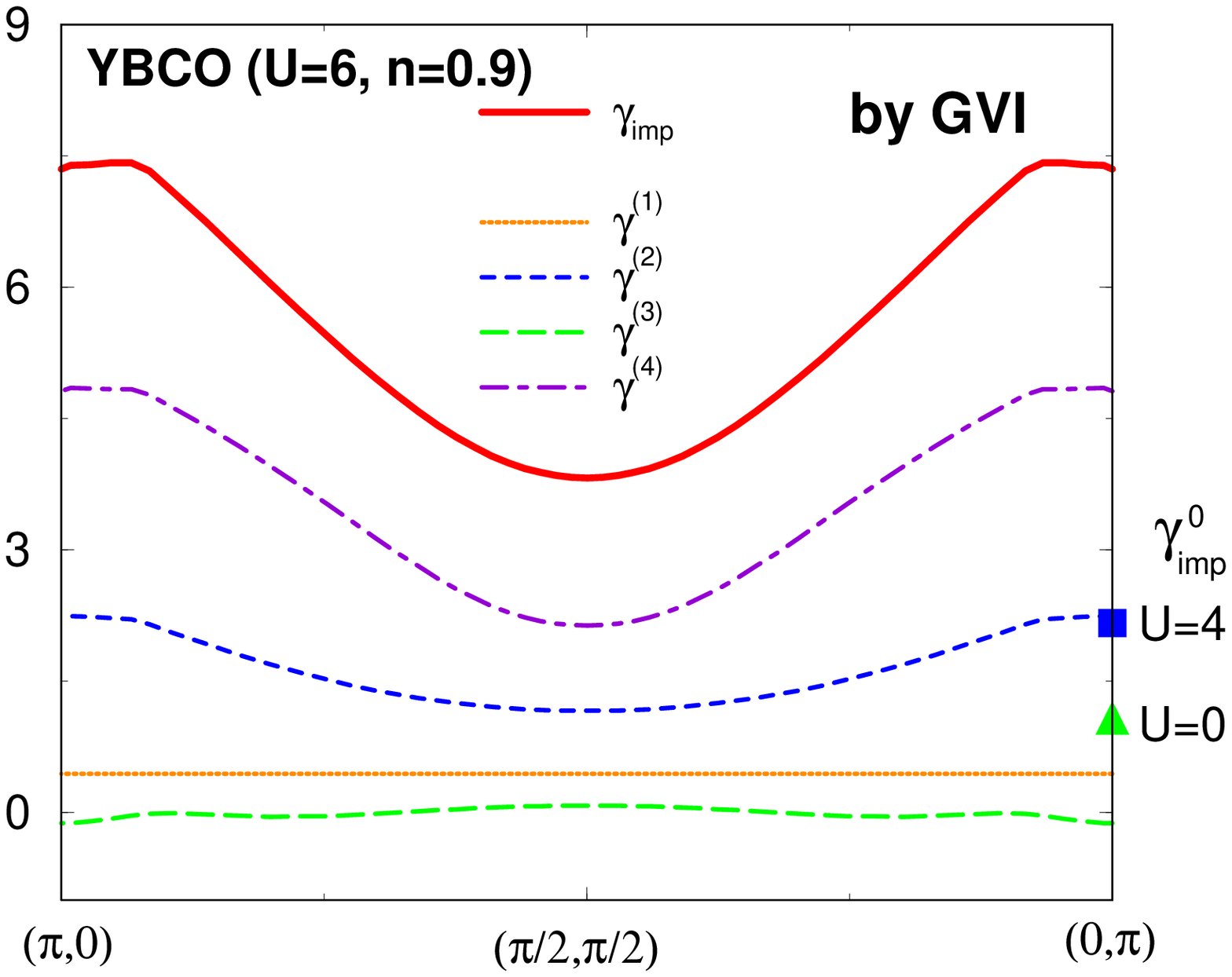,width=6.5cm}
\end{center}
\caption{(Color online) 
$\gamma_\k^{\rm imp(l)}$ ($l=1\sim4$)
per $n_{\rm imp}$ at $T=0.02$, along the Fermi line.
They are given by the $GV^I$-method.
They represent contributions by 
eqs.(\ref{eqn:T1})-(\ref{eqn:T4}), respectively.
}
\label{fig:Gamma-1234}
\end{figure}

Here, we examine the origin of the enhancement of 
$\gamma_\k^{\rm imp}$ in more detail.
Figure \ref{fig:Gamma-1234} shows 
$\gamma_\k^{\rm imp(1)}$ ($l=1\sim4$)
for LSCO and YBCO, which represent contributions by 
eqs.(\ref{eqn:T1})-(\ref{eqn:T4}), respectively.
They are also expressed in Fig. \ref{fig:Tmatrix}.
Note that $\gamma_\k^{\rm imp}= \sum_{l=1}^4 
\gamma_\k^{\rm imp(l)}$.
In both YBCO and LSCO, $\gamma_\k^{\rm imp(2)}$ and 
$\gamma_\k^{\rm imp(4)}$ give main contributions:
The latter is dominant for YBCO, whereas 
$\gamma_\k^{\rm imp(2)}$ is comparable to $\gamma_\k^{\rm imp(4)}$
around $\k=(\pi/2,\pi/2)$ for LSCO.
In both systems, $\gamma_\k^{\rm imp(4)}$ grows
drastically below $T=0.02$ as $U$ is increased.
In the same way, $\gamma_\k^{\rm imp(4)}$ 
takes a large value for NCCO at lower temperatures.
This enhancement of $\gamma_\k^{\rm imp(4)}$
at lower temperatures gives rise to the insulating
behavior of the resistivity, as we will show later.

\subsection{Resistivity}
\label{sec:rho}

Here, we calculate the resistivity in nearly AF metals
in the presence of strong impurities.
Hereafter, we take account of the impurity effect
only up to $O(n_{\rm imp})$.
In other words, we neglect the interference effect 
(e.g., the weak localization effect) which is given by the 
higher order terms with respect to $n_{\rm imp}$.
In fact, many anomalous impurity effects of HTSC
in which we are interested are of the order of $O(n_{\rm imp})$.
For example, the residual resistivity in HTSC is proportional
to the impurity concentration for $n_{\rm imp}\simle 4$\% \cite{Uchida}.
Surprisingly, the residual resistivity per impurity 
increases drastically as the system approaches the half-filling.
The relation $\Delta\rho \sim (4\hbar/e^2)n_{\rm imp}/\delta$ 
holds in the under-doped region, which is $n/\delta$ times larger than 
the residual resistivity in 2D electron gas.
($\delta=|1-n|$ is the carrier doping concentration.)
Hereafter, we will explain this experimental fact
based on the idea that the effective cross section of an impurity 
is enlarged due to many body effect near the AF-QCP.

\begin{figure}
\begin{center}
\epsfig{file=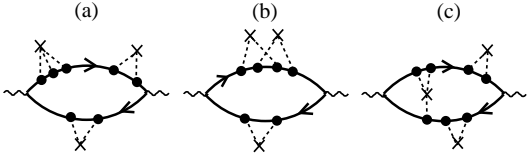,width=7cm}
\end{center}
\caption{Several diagrams for the conductivity
in the presence of impurities.
The residual resistivity $\Delta\rho$ given by type (a) and (c) 
diagrams is proportional to $O(n_{\rm imp})$.
On the other hand, $\Delta\rho$ given by type (b) diagrams 
is $O(n_{\rm imp}^2)$.
In the present study, we take only type (a) diagrams into account.
}
\label{fig:Rho-diagram}
\end{figure}

The conductivity is given by the two-particle Green function;
we show some diagrams in Fig. \ref{fig:Rho-diagram}.
Here, the cross represents the impurity potential,
and the filled circle is the three point vertex
due to the electron-electron correlation.
In Fig. \ref{fig:Rho-diagram}, the type (a) diagrams give the correction
of the order of $O(n_{\rm imp})$, whereas the type (b) diagrams
are $O(n_{\rm imp}^2)$ 
because they contain cross terms between different impurities
 \cite{Langer}.
Thus, we drop all the diagrams which contain cross terms.
The type (c) diagrams, which contain current vertex corrections (CVC's)
due to impurities, which are necessary to cancel the effect of
forward scattering on the conductivity.
However, we drop all the diagrams with CVC's
for the simplicity of the calculations,
which would be allowed for a qualitative discussion.
We also drop the CVC due to the electron-electron correlation
because it gives merely a small correction to the conductivity,
as shown in ref. \cite{Kontani-Hall}.

As a result, by neglecting the CVC's,
the conductivity without and with impurities up to $O(n_{\rm imp})$, 
$\sigma_0$ and $\sigma_{\rm imp}$, are given by the following equations
 \cite{Eliashberg,Langer,comment}:
\begin{eqnarray}
\sigma_0&=& e^2\sum_\k \int\frac{d\e}{\pi}
 \left( -\frac{\d f}{\d\e} \right)
|G_\k^0(\e)|^2 v_{\k x}^2(\e) ,
 \\
\sigma_{\rm imp}&=& e^2\sum_\k \int\frac{d\e}{\pi}
 \left( -\frac{\d f}{\d\e} \right)
|{\bar G}_\k(\e)|^2 v_{\k x}^2(\e) ,
 \label{eqn:sigma-imp}
\end{eqnarray}
where $v_{\k x}(\e)= \d\e_\k/\d k_x + {\rm Re}\Sigma_\k(\e)$.
$G_\k^0(\e)= (\e+\mu-\e_\k-\Sigma_\k(\e))^{-1}$
is the Green function obtained by the FLEX approximation
without impurity.
The averaged Green function in the presence of impurities,
${\bar G}_\k$, is given by
\begin{eqnarray}
{\bar G}_\k(\e-i\delta)=  \left( 
 \{G_\k^0(\e-i\delta)\}^{-1}-i\gamma_\k^{\rm imp}(\e) \right)^{-1} .
 \label{eqn:Gbar}
\end{eqnarray}
%

\begin{figure}
\begin{center}
\epsfig{file=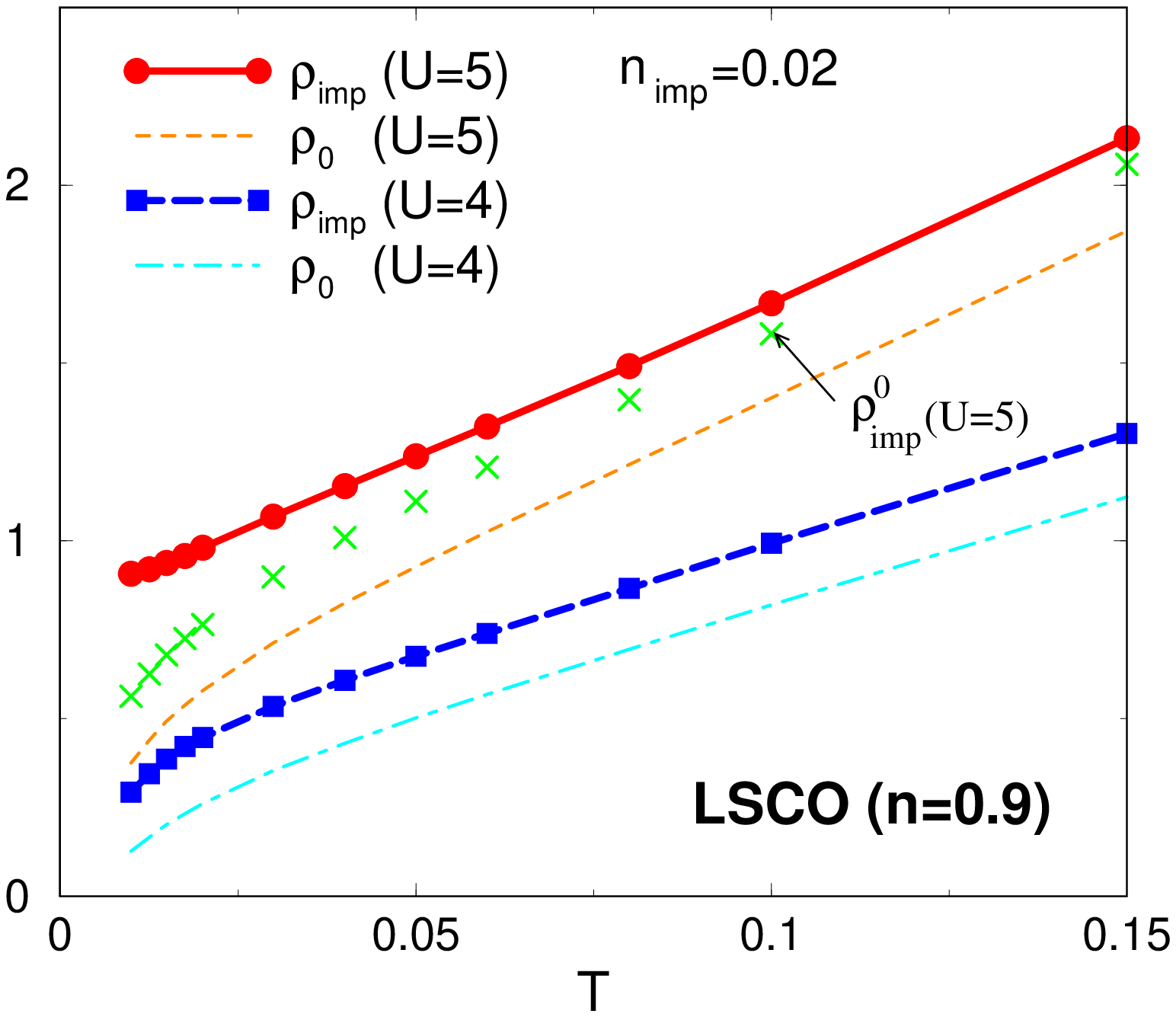,width=7.5cm}
\epsfig{file=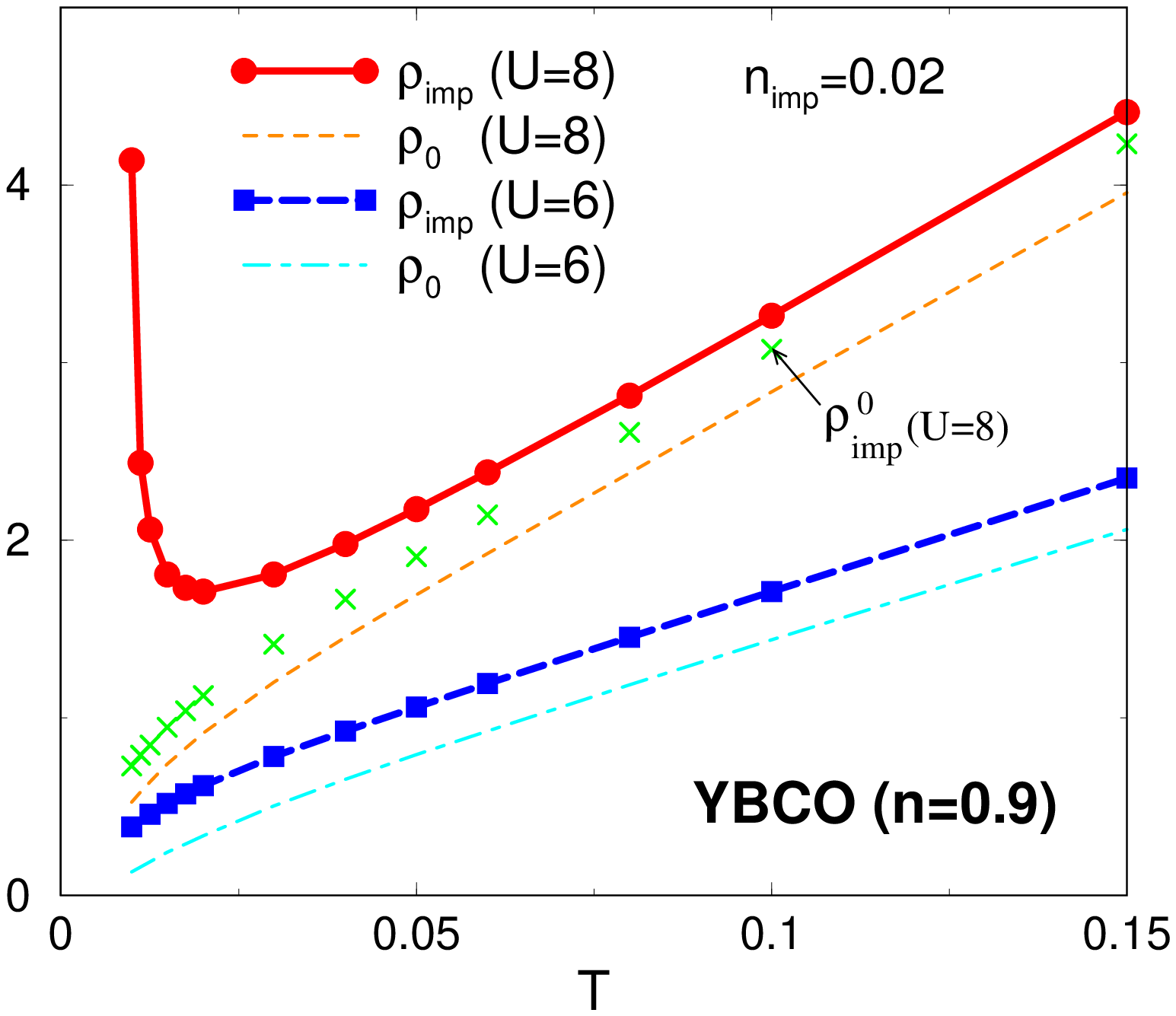,width=7.5cm}
\epsfig{file=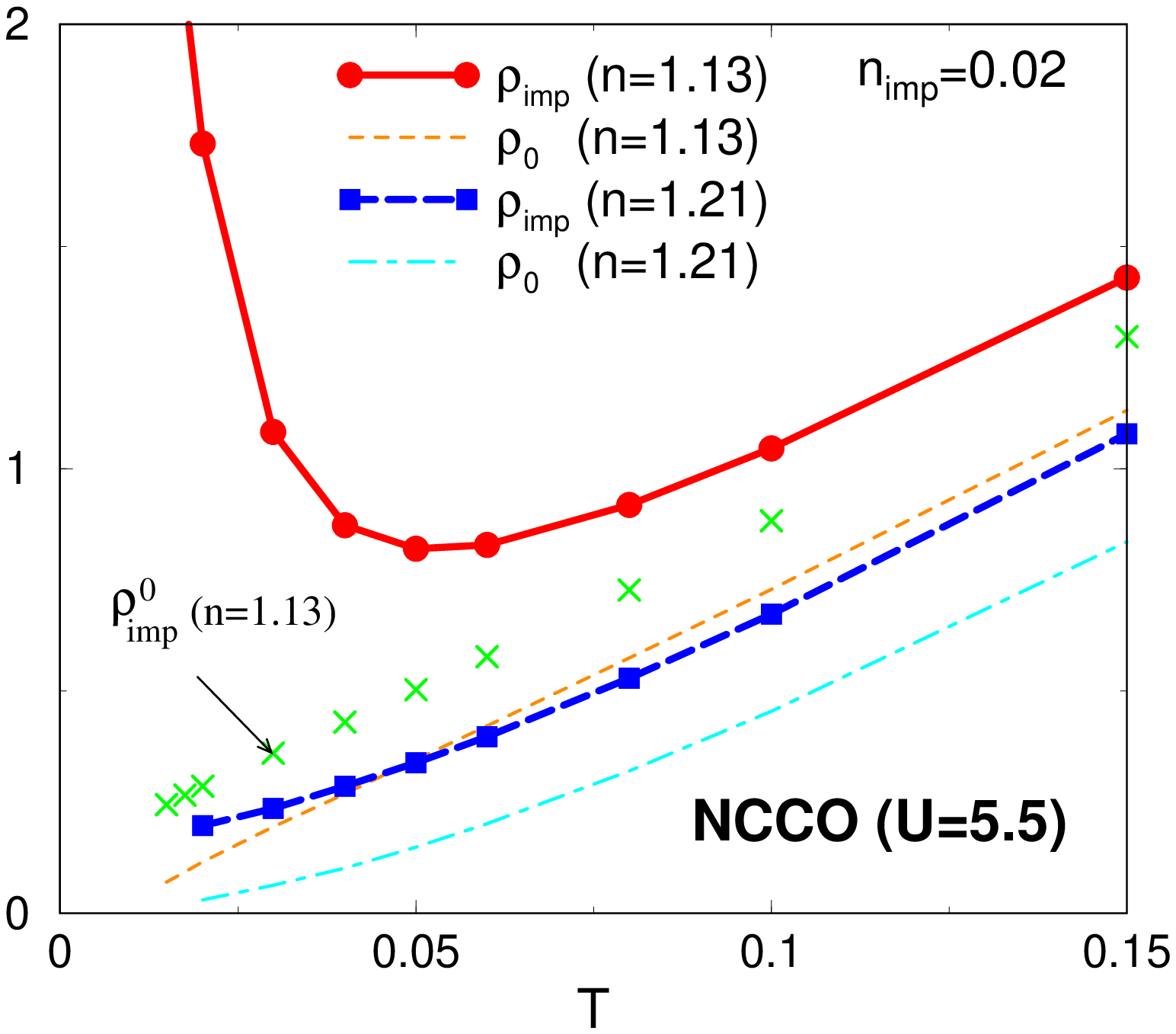,width=7.5cm}
\end{center}
\caption{(Color online) 
$\rho_{\rm imp}$ ($\rho_{\rm imp}^0$) 
represents the resistivity for $n_{\rm imp}=0.02$ 
given by the $GV^I$-method (local scattering approximation).
$\rho_0$ is the resistivity for the host system.
At higher temperatures, the residual resistivity 
$\rho_{\rm imp}-\rho_0$ is almost $T$-independent, 
and its value increases near the AF-QCP.
At lower temperatures, the insulating behavior emerges
in the close vicinity of the AF-QCP.
$T=0.1$ and $\rho=1$ correspond to 400K and 250$\mu\Omega\cdot$cm
for HTSC's, respectively.
}
\label{fig:rho}
\end{figure}

Figure \ref{fig:rho} shows the
temperature dependences of the resistivities
for LSCO, YBCO and NCCO obtained by the $GV^I$-method;
$\rho_{\rm imp}=1/\s_{\rm imp}$ for $n_{\rm imp}=0.02$.
$\rho_0=1/\s_0$ is the resistivity without impurity.
We see that the ``residual resistivity'' at finite $T$, 
which is the increment of resistivity due to impurity
$\Delta\rho\equiv \rho_{\rm imp}-\rho_0$,
is approximately constant for a wide range of $T$,
except for the abrupt increase at lower temperatures in YBCO and NCCO.
$\Delta\rho$ grows as we enlarge $U$ (YBCO, LSCO)
or change the filling number $n$ towards the half-filling (NCCO).
This result comes from the fact that 
the scattering cross section of an impurity is enlarged
by the nonlocal modulation of the self-energy, $\delta\Sigma$.
In other words, non s-wave (elastic and inelastic) 
scatterings caused by $\delta\Sigma$ gives an anomalously
large residual resistivity near the AF-QCP.

Figure \ref{fig:rho} also shows that
$\rho_{\rm imp}^0$ derived from eq.(\ref{eqn:sigma-imp})
by replacing $\gamma_\k^{\rm imp}(\e)$ in eq.(\ref{eqn:Gbar})
with $\gamma_{\rm imp}^0(\e)$ which is momentum independent.
We call it the ``local scattering approximation''
because non-s-wave impurity scattering processes caused by 
the nonlocal effect ($\delta\Sigma$) are dropped.
In contrast to $\Delta\rho$,
the doping dependences of 
$\Delta\rho^0\equiv \rho_{\rm imp}^0-\rho_0$ is much moderate.
Moreover, $\Delta\rho^0$ slightly decreases at low temperatures
as $\gamma_\k$ becomes anisotropic:
In fact,
$\displaystyle \langle \frac1{\gamma_\k+\gamma_{\rm imp}^0}
 \rangle^{-1}- \langle \frac1{\gamma_\k} \rangle^{-1} \ll 
\gamma_{\rm imp}^0$ when $\gamma_\k (\gg \gamma_{\rm imp}^0)$
is very anisotropic.

In each compound (YBCO, LSCO, NCCO), the average
spacing between CuO$_2$-layers is about 6\AA.
Using the relation $h/e^2=26$k$\Omega$,
$\rho=1$ in the present calculation corresponds to 
250$\mu\Omega\cdot$cm.
As shown in fig. \ref{fig:rho},
residual resistivities for $n_{\rm imp}=0.02$ at $T=0.05$ 
for LSCO ($U=5$), YBCO ($U=8$) and NCCO ($n=1.13$) are 
0.7, 1.0 and 0.8, respectively.
They correspond to $175\sim 250\mu\Omega\cdot$cm.
These obtained values are close to the
experimental residual resistivity in under-doped HTSC's;
$\Delta\rho\sim (4\hbar/e^2)n_{\rm imp}/|1-n|$
 \cite{Uchida}.

Another important findings is the insulating behavior 
$(d\rho_{\rm imp}/dT<0)$ in YBCO and NCCO at lower temperatures.
We also checked that a similar upturn is also observed 
in LSCO for $U=6$.
This insulating behavior of $\rho_{\rm imp}$ is caused by the 
steep increment of $\gamma_\k^{\rm imp}$ at lower temperatures, 
which is mainly given by 
$\gamma_\k^{\rm imp(4)} \propto {\rm Im}\delta{\hat \Sigma}(0)$.
Therefore, the physical origin of this phenomenon is the
strong inelastic scattering around the impurity.
The obtained insulating behavior will be universal
for systems with strong nonmagnetic impurities
in the close vicinity of the AF-QCP.
Actually, the upturn of $\rho$ is widely observed in under-doped HTSC's,
both in hole-doped and electron-doped compounds
\cite{Uchida,Ando1,Ando2,Sekitani,Shibauchi}.
The origin is not the weak localization
as we explained in \S \ref{sec:intro}.
Based on the present study,
we expect that residual disorder will be the origin of the 
insulating behavior in HTSC's.

\begin{figure}
\begin{center}
\epsfig{file=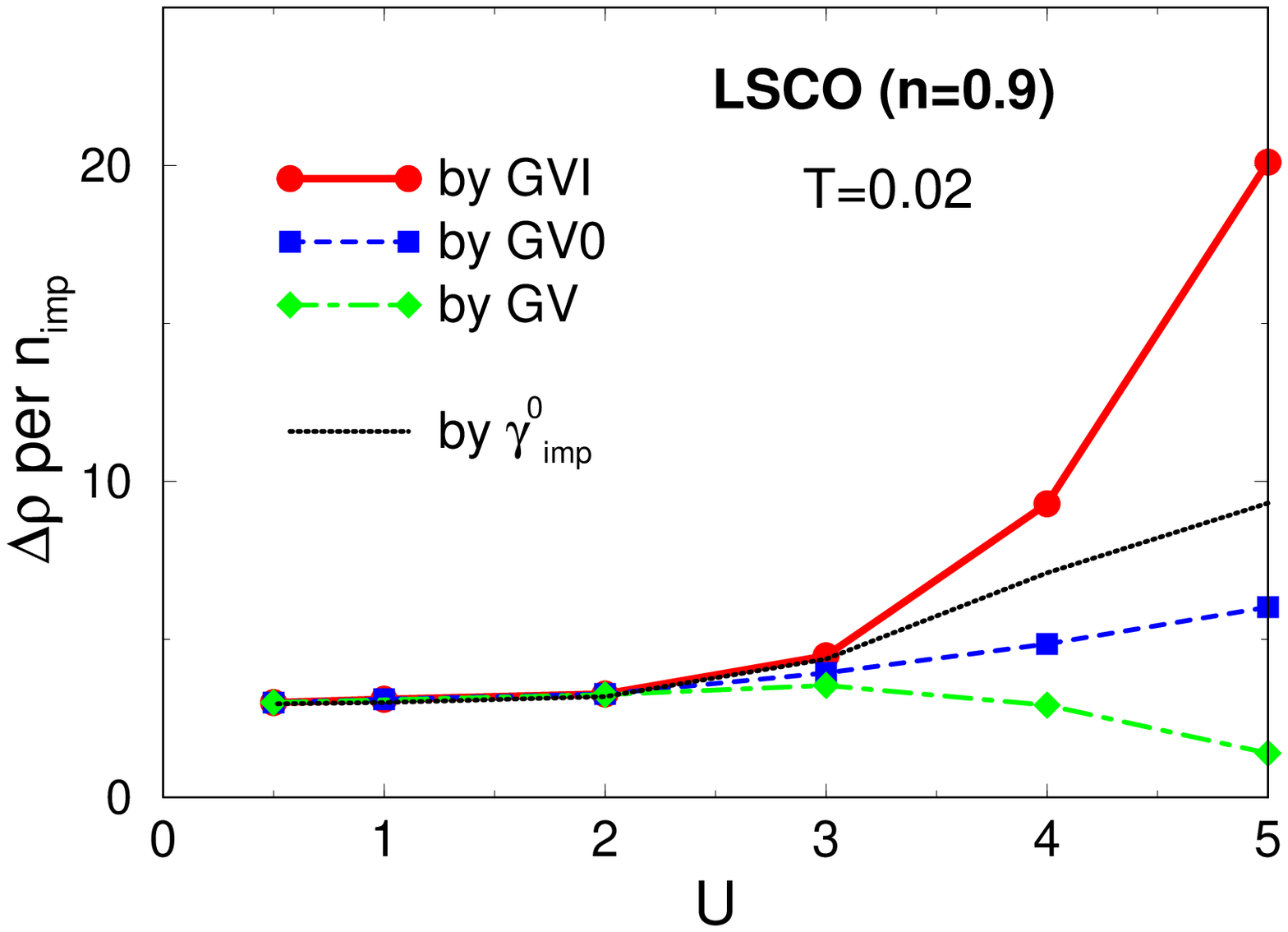,width=7.5cm}
\epsfig{file=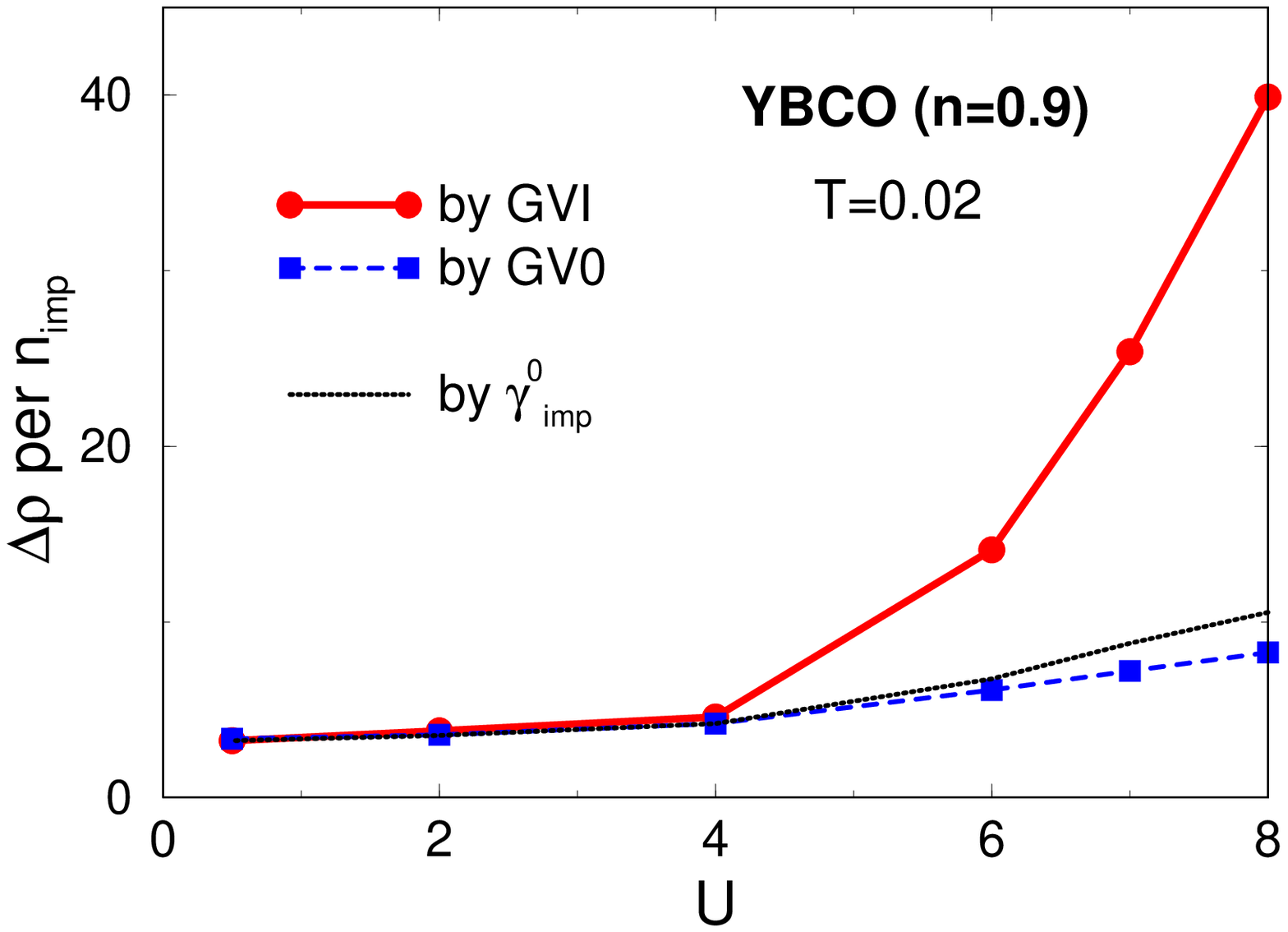,width=7.5cm}
\end{center}
\caption{(Color online) 
$\Delta\rho \equiv \rho_{\rm imp}-\rho_0$
per $n_{\rm imp}$ at $T=0.02$.
$\Delta\rho$ given by the $GV^I$-method grows drastically 
as $U$ increases, whereas $\Delta\rho$ by the $GV$-method
gradually drops.
Here, ``by $\gamma_{\rm imp}^0$'' represents the
residual resistivity by the local approximation,
which slightly increases with $U$ in proportion to
$\gamma_{\rm imp}^0 \propto 1/N_{\rm host}(0)$.
}
\label{fig:rho-T002}
\end{figure}

In Fig. \ref{fig:rho-T002}, 
we show that $U$ dependences of $\Delta\rho$ and $\Delta\rho^0$
(per impurity) at $T=0.02$ given by the $GV^I$-method,
both for LSCO and YBCO.
$\Delta\rho^0$ increases as $U$ is raised, inversely proportional 
to the DOS given by the FLEX approximation, $N_{\rm host}(0)$.
We stress that $\Delta\rho$ increases drastically as $U$ is raised,
much faster than $\Delta\rho^0$ does.
This prominent increment is derived from the strong inelastic 
scattering around the impurity, Im$\delta{\hat \Sigma}(0)$.
For comparison, we also show results
given by the $GV$ and $GV^0$-methods in Fig. \ref{fig:rho-T002}.
$\Delta\rho_{\rm [GV0]}$ is close to $\Delta\rho^0$
because the enhancement of AF fluctuations around the impurity
are not taken into account.
Even worse, $\Delta\rho_{\rm [GV]}$ decreases for $U>3$ in LSCO.
This result should be an artifact of the $GV$-method,
because it fails to reproduce the enhancement of spin fluctuations
as explained in previous sections.

The prominent $U$-dependence of $\Delta\rho$ given by $GV^I$-method
would corresponds to the pressure dependence of $\Delta\rho$
observed in $\kappa$-(BEDT-TTF)$_4$Hg$_{2.89}$Br$_8$,
which stays near the AF-QCP at ambient pressure
 \cite{Taniguchi}.
In this compound, $\Delta\rho$ decreases to one tenth of its 
original value by applying the pressure.
Considering that the applied pressure makes $U/W_{\rm band}$ small,
this experimental result is consistent with Fig. \ref{fig:rho-T002}.
We also note that the residual resistivity 
in heavy fermion systems near the magnetic instabilities 
frequently show prominent pressure dependence;
$\Delta\rho$ takes a maximum value at the AF-QCP,
and it decreases quickly as the system goes away from the AF-QCP.
 \cite{Jaccard1,Jaccard2,CeCuAu,Ce115,Ce115-2}.
This experimental fact is well explained by Fig. \ref{fig:rho-T002}.

\begin{figure}
\begin{center}
\epsfig{file=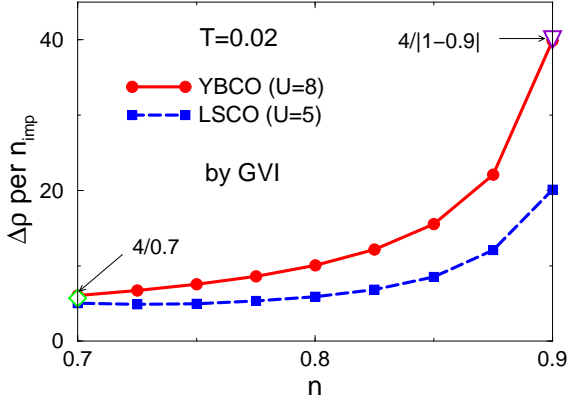,width=7.5cm}
\end{center}
\caption{(Color online) 
Filling-dependence of $\Delta\rho/n_{\rm imp}$ 
given by the $GV^I$-method at $T=0.02$.
Origin of the huge $\Delta\rho$ in under-doped region
is the short $\tau$ around impurities
induced by strong AF fluctuations.
}
\label{fig:rho-n}
\end{figure}

Figure \ref{fig:rho-n}
shows the filling dependence of $\Delta\rho$
given by the $GV^I$-method at $T=0.02$, which is above the
upturn temperature of $\rho$ for YBCO as recognized in Fig. \ref{fig:rho}.
In both LSCO and YBCO, 
$\Delta\rho$ increases drastically as $n$ approaches unity,
far beyond the s-wave unitary scattering value ($\sim 4/n$).
The obtained result is consistent with experimental observations in HTSC's,
where $\Delta\rho \sim \Delta\rho^0$ in the over-doped region, 
whereas $\Delta\rho \gg \Delta\rho^0$ in the under-doped region
  \cite{Uchida}.

In the present calculation, we dropped all the CVC's for simplicity.
As shown in ref. \cite{Kontani-Hall},
$\rho_0$ without impurities is slightly enlarged by the CVC;
effect of the CVC for $\rho_0$ is small because the origin of
the resistivity is the large angle scattering due to 
AF fluctuations($\q \approx (\pi,\pi)$).
[In general, CVC's for the resistivity are important 
when small angle scatterings are dominant.]
In the present calculation,
the origin of the huge $\Delta\rho \propto \gamma_\k^{\rm imp(4)}$
is also the AF fluctuations induced around the impurity.
Therefore, we expect that obtained $\Delta\rho$ in this section
is qualitatively reliable.
On the other hand, CVC's play quite important roles 
on $R_{\rm H}$, $\Delta\rho/\rho$ and $\nu$
 \cite{Kontani-rev}:
It is an important future issue to study these transport
coefficients in the presence of impurities
by taking CVC's into account.
Finally, 
we note that the obtained insulating behavior of $\rho_{\rm imp}$
might be under-estimated because the area of region A
in the present numerical study ($M=6\sim8$) would be 
not enough in the very close vicinity of the AF-QCP 
($\xi_{\rm AF}\gg 1$).

\subsection{Decrease of the Hole Density $n_{\rm h}$ ($=|1-n|$)
around the Impurity, and Increase of $\Delta\rho$ at $T=0$}
\label{sec:rho0}

Up to now, we have found that 
$\gamma_\k^{\rm imp(4)}$ (and $\gamma_\k^{\rm imp(2)}$)
give the main contribution to the huge residual resistivity
$\Delta\rho_{\rm imp}$ at finite temperatures.
However, $\gamma_\k^{\rm imp(4)}(0)=0$ at zero temperature
because $\delta\Sigma_{l,m}(0)$ becomes a real function at $T=0$.
Therefore, it is highly nontrivial to predict the value of
$\Delta\rho_{\rm imp}$ at $T=0$.
On the other hand, $\gamma_\k^{\rm imp(l)}$, 
$\gamma_\k^{\rm imp(2)}$ and $\gamma_\k^{\rm imp(3)}$
could give an enlarged $\Delta\rho_{\rm imp} \gg \Delta\rho_{\rm imp}^0$ 
at zero temperatures, because they are finite even at $T=0$.

Hereafter, we discuss the residual resistivity at $T=0$.
For simplicity, we assume an two-dimensional 
isotropic system with the dispersion $\e_k= \k^2/2m$.
At $T=0$, $\gamma_\k^{\rm imp}$ is given by
\begin{eqnarray}
\gamma_\k^{\rm imp}&=& n_{\rm imp}{\rm Im}T_{\k,\k}
 \nonumber \\
&=&  n_{\rm imp} \frac{2}{m}\sum_l \sin^2\delta_l ,
\end{eqnarray}
where $l$ represents the angular momentum with respect to the 
impurity potential ($l=0,\pm1,\pm2,\cdots$), 
and $\delta_l$ is the phase shift for channel $l$.
If we drop the CVC, the resistivity at $T=0$ is given by
 \cite{Langer}
\begin{eqnarray}
\rho_{\rm imp}=\frac{4\hbar n_{\rm imp}}{e^2 n} \sum_l \sin^2\delta_l .
\end{eqnarray}
Here, $\sin^2\delta_l$ is replaced with
$\sin^2\delta_l - \cos(\delta_l-\delta_{l+1})\sin\delta_l\sin\delta_{l+1}$
if the CVC due to impurity is taken into account.
On the other hand,
the number of localized electrons around the impurity, $\Delta n_{\rm tot}$,
in an isotropic 2D system is given by
 \cite{sumrule}
\begin{eqnarray}
\Delta n_{\rm tot}= \frac{2}{\pi}\sum_l \delta_l .
\end{eqnarray}
Thus, the residual resistivity at $T=0$ will grows
as the number of electrons in bound states increases.

\begin{table}[hbt]
YBCO 
\begin{center}
\begin{tabular}{|c||l|l|l|l|}
\hline
$T$ & 0.01 \ \ & 0.015\ & 0.02 \ \ & 0.05 \ \ \\
\hline \hline
$100\times\Delta n(1,0)$ & 6.15 & 1.50 & 1.06 & 0.60 \\
\hline
$\Delta n_{\rm tot}$ & 2.86 & 0.334 & 0.190 & 0.052 \\
\hline
$\a_{\rm St}$ & 0.9968 & 0.9950 & 0.9932 & 0.9807 \\
\hline
\end{tabular}
\end{center}
NCCO 
\begin{center}
\begin{tabular}{|c||l|l|l|l|}
\hline
$T$ & 0.02 \ \ & 0.025\ & 0.03 \ \ & 0.06 \ \ \\
\hline \hline
$100\times\Delta n(1,0)$ & -3.84 & -3.12 & -2.78 & -2.13 \\
\hline
$\Delta n_{\rm tot}$ & -1.52 & -0.897 & -0.639 & -0.254 \\
\hline
$\a_{\rm St}$ & 0.9965 & 0.9958 & 0.9948 & 0.9839 \\
\hline
\end{tabular}
\end{center}
\caption{Temperature dependences of $\Delta n(1,0)$,
$\Delta n_{\rm tot}$ and the Stoner factor 
$\a_{\rm St}={\rm max}_\q U\Pi_\q^0(0)$.
}
\label{table1}
\end{table}
Table \ref{table1} represent $\Delta n(1,0) \equiv n(1,0)- n$
and  $\Delta n_{\rm tot} \equiv \sum_{\bf r} \Delta n({\bf r})$
for both YBCO ($n=0.9$, $U=8$) and NCCO ($n=1.13$, $U=5.5$). 
We see that the number of electron at ${\bf r}=(1,0)$
approaches the half filling ($n=1$) at low temperatures
for both YBCO and NCCO.
Moreover, $\Delta n_{\rm tot}$ increases (decreases) prominently
in YBCO (NCCO) as $T$ decreases.
This result suggests that the residual resistivities at $T=0$
will be larger than $\rho_{\rm imp}^0$ in the vicinity of the AF-QCP.

Finally, we discuss why $n(1,0)$ approaches unity 
and $\Delta n_{\rm tot}$ increases (decreases) in hole-doped
(electron-doped) systems.
In the FLEX approximation, the thermodynamic potential $\Omega$
in a uniform system is given by \cite{Ikeda}
\begin{eqnarray}
\Omega&=& -T\sum_{\q,l}{\rm Tr}\left[{\Sigma}{G}
 + {\rm ln}(-[{G}^0]^{-1}+{\Sigma}) \right]
 \nonumber \\
& &+ T\sum_{\q,l} {\rm Tr}\left[ \frac32 {\rm ln}(1-U{\Pi^0})
 + \frac12 {\rm ln}(1+U{\Pi^0}) \right.
 \nonumber \\
& & \ \ \left. + U {\Pi^0}+U^2 [{\Pi^0}]^2 \right] .
 \label{eqn:Omega}
\end{eqnarray}
As we explained, the AF fluctuations are enhanced around the impurity.
This effect would be expressed by increasing
$\Pi^0(\q,\w_l)$ in eq. (\ref{eqn:Omega}) by $\Delta\Pi^0$ $(>0)$, 
or $U$ by $\Delta U$ $(>0)$.
According to eq. (\ref{eqn:Omega}),
\begin{eqnarray}
\frac{\d\Omega}{\d U}&\approx& 
 -\frac{3}{2} T\sum_{\q,l} {\Pi^0}(\q,\w_l)
 \left( 1-U{\Pi^0}(\q,\w_l) \right)^{-1} ,
 \label{eqn:dOdU}
\end{eqnarray}
in the case of $1-U{\Pi^0}(\Q,0)\ll 1$.
In deriving (\ref{eqn:dOdU}), we used the fact that 
the implicit derivative through $\Sigma$ vanishes because of 
the stationary condition $\delta \Omega/ \delta \Sigma=0$
in the conserving approximation \cite{Bickers,Baym-Kadanoff}.
According to eq.(\ref{eqn:dOdU}), we obtain
\begin{eqnarray}
\frac{\d n}{\d U}
&=& -\frac{\d}{\d U}\left( \frac{\d \Omega}{\d \mu} \right)
 \nonumber \\
&=& \frac32 T\sum_{\q,l} \frac1{(1-U\Pi^0(\q,\w_l))^{2}}
 \frac{\d \Pi^0(\q,\w_l)}{\d\mu} .
 \label{eqn:dndU}
\end{eqnarray}
As a result,
the electron density $n$ around the impurity, 
where $\Delta U>0$ is satisfied as mentioned above,
increase (decreases) when $\d\Pi(\Q,0)/\d\mu$ is positive (negative).
Therefore, we conclude that $\Delta n_{\rm tot}$ 
will increase (decrease) in hole-doped (electron-doped) systems,
as recognized by numerical results given by the $GV^I$-method.


\section{Discussions}
\label{sec:disc}

\subsection{Summary of the Present Work and Future Problems}
\label{sec:sum}
The present study reveals that a single impurity 
strongly affects the electronic states in a wide area around the impurity
in the vicinity of the AF-QCP.
For this purpose, we developed the $GV^I$-FLEX method, which is a 
powerful method to study the impurity effect in strongly correlated systems.
The $GV^I$ method is much superior to the $GV$, which is a
fully self-consistent FLEX approximation.
Using the $GV^I$ method, characteristic impurity effects in under-doped 
HTSC's are well explained in a unified way, without introducing
any exotic mechanisms assuming the breakdown of the Fermi liquid state.
The main numerical results 
are shown in Fig. \ref{fig:DOS2D} (local DOS around the impurity site),
Figs. \ref{fig:kaiS}-\ref{fig:kaiS-AF} (local and staggered 
susceptibilities), and Figs. \ref{fig:rho}-\ref{fig:rho-n}
(resistivity in the presence of impurities).
Qualitatively, these obtained numerical results are very similar
for YBCO, LSCO and NCCO.  
Therefore, novel impurity effects in nearly AF metals
revealed by the present work would be be universal.

\begin{figure}
\begin{center}
\epsfig{file=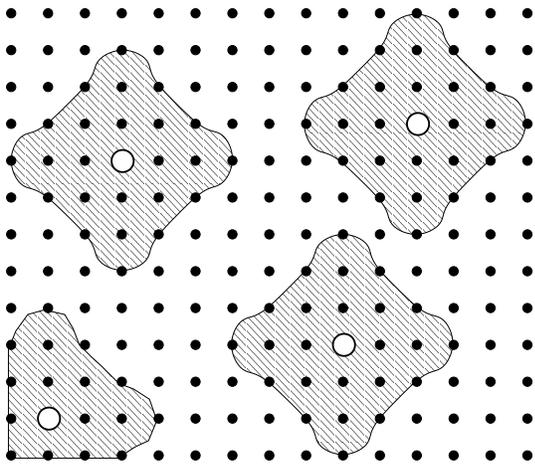,width=7cm}
\end{center}
\caption{Swiss cheese structure induced by nonmagnetic
impurities ($n_{\rm imp}= 0.021$) denoted by open circles,
which replace Cu sites randomly.
In the shadow area, both local and staggered spin susceptibilities
are enhanced, and the quasiparticle lifetime is depressed.
The radius of the Swiss cheese hole is $\sim \xi_{\rm AF}$.
}
\label{fig:swiss}
\end{figure}

Based on the $GV^I$ method,
we found that both local and staggered susceptibilities
are prominently enhanced around the impurity site,
as shown in Figs. \ref{fig:kaiS} and \ref{fig:kaiS}.
Especially, a nonmagnetic impurity causes a Curie-like 
spin susceptibility, $\mu_{\rm eff}^2/3T$. 
The $GV^I$-method gives $\mu_{\rm eff}\approx 0.74\mu_{\rm B}$
for LSCO ($n=0.9$, $U=5$), shown in Figs. \ref{fig:kaiS-AF}.
Note that $\mu_{\rm eff} \sim 1\mu_{\rm B}$
in YBa$_2$Cu$_3$O$_{6.66}$ ($T_{\rm c}\approx60K$).
We also found that 
the quasiparticle damping rate takes a huge value
around the impurity, owing to the enhanced AF fluctuations.
By this reason, the local DOS at the Fermi level
is strongly suppressed around the impurity site,
which forms a so-called ``Swiss cheese structure''
as shown in Fig. \ref{fig:swiss}.
Its radius is about the AF correlation length of the host, 
$\xi_{\rm AF}$, which is about $3\sim4$a (a being the lattice spacing)
in slightly under-doped HTSC's at $T=0.02$.
We guess that Swiss cheese holes stay in a normal state even below 
$T_{\rm c}$, because of the extremely short quasiparticle lifetime there.
In fact, the residual specific heat ($T\ll T_{\rm c}$) 
induced by an impurity becomes very large in under-doped systems 
 \cite{Ido2}.
This experimental fact will be explained in our future study
based on the $GV^I$-method
 \cite{future}.

Near the AF-QCP, the short quasiparticle lifetime
inside the Swiss cheese hole gives rise to 
a huge residual resistivity $\Delta\rho$,
as shown in Figs. \ref{fig:rho}-\ref{fig:rho-n}.
In the under-doped region,
$\Delta\rho$ grows far beyond the s-wave unitary scattering limit
$\sim (4\hbar/e^2)n_{\rm imp}/n$.
We find that $\Delta\rho$ is almost $T$-independent
for a wide range of temperature, and it increases drastically
as the system approaches the AF-QCP.
This result is consistent with experiments for HTSC's.
The obtained value of $\Delta\rho$, 
$175\sim250\mu\Omega\cdot$cm for $n_{\rm imp}=0.02$,
are recognized in under-doped HTSC's
 \cite{Uchida}.
Furthermore, in the close vicinity of the AF-QCP, the resistivity 
given by the $GV^I$-method 
shows the ``Kondo-like`` insulating behavior ($d\rho/dT<0$) 
under the presence of 
nonmagnetic impurities with low concentration ($\sim2$\%).
This surprising result would explain the ``upturn of resistivity''
which is frequently observed in under-dope HTSC's,
by assuming the existence of residual disorders.
The mechanism of this insulating behavior had been a long-standing 
unsolved issue in under-doped HTSC's.
Different from a conventional single-channel Kondo effect,
the residual resistivity given by the $GV^I$-method 
grows far beyond $(4\hbar/e^2)n_{\rm imp}/n$.

We stress that that $\gamma_\k^{\rm imp}$ given by the $GV^I$-method
has strong $\k$-dependence, so the structure of ``hot/cold spots'' 
is maintained against impurity doping.
This result could be examined by the ARPES measurements.
We also comment that the CVC's due to spin fluctuations cause 
a finite ``residual resistivity'', if we define it as the 
extrapolated value at $T=0$, even in the absence of impurities 
\cite{Kontani-Hall}.
We have to take this fact into account in analysing experimental data.

In the FLEX approximation, the AF-order (in the RPA) 
is suppressed by thermal and quantum fluctuations, 
which is expressed by $\Sigma^0$.
In the $GV^I$-method, the reduction of fluctuations around the 
impurity site gives rise to the enhancement of susceptibility.
However, this mechanism is absent in the RPA.
Therefore, the enhancement of susceptibility is tiny within the RPA.
In addition, in the $GV^I$-method,
the spin (charge) susceptibility $\chi^{\rm Is(c)}$ 
contains the self-energy correction by $U$'s
and that by $I$'s are treated on the same footing,
whereas the cross term $\delta\Sigma$ is dropped 
because it will be cancelled out by VC's in large part.
On the other hand, $\chi^{\rm RPA}$ given by the RPA
contains only self-energy correction by $I$'s
 \cite{Bulut01,Bulut00,Ohashi}.
This equal footing treatment of the correlation effect 
and the impurity effect in the $GV^I$-method would be 
the reason for the superiority of this method.

In future, we will study the transport phenomena
by taking the current vertex corrections (CVC) accurately,
in order to solve the impurity effect on various
transport coefficients in HTSC's and in related systems.
 \cite{future}.
We will also study the superconducting state around 
a nonmagnetic impurity in under-doped HTSC's 
 \cite{future},
to explain experimental observations given by STM/STS measurements
 \cite{Pan,Davis,Ido}.
As for HF systems and in organic metals, the strength of 
residual impurity (or disorder) potential would be
comparable with bandwidth $W_{\rm band}$.
Therefore, we have to study to what extent the obtained results
in the present paper (for $I=\infty$) hold in the case of 
$I\sim W_{\rm band}$.

\subsection{Possibility of the Impurity-Induced Magnetic Order}
\label{sec:comparison}
In the $GV^I$-method, the self-energy for the host system 
is given by the FLEX approximation,
which cannot be applicable for heavily under-doped systems
near the Mott insulating state.
However, FLEX approximation gives qualitatively reliable 
results for slightly under-doped region ($|1-n|\sim 0.1$)
to over-doped region
 \cite{Takimoto,Koikegami,Wermbter,Kontani-Hall,Manske-PRB}.
To obtain quantitatively reliable results for under-doped region,
vertex correction (VC) for the self-energy will be necessary,
as indicated by ref. \cite{Schmalian}.
The pseudo-gap phenomena under $T^\ast$ are well reproduced
by the FLEX+$T$-matrix approximation, where self-energy correction 
due to strong superconducting (SC) fluctuations are taken into account
 \cite{Yamada-rev,Dahm-T,Kontani-rev,Kontani-N}.
By taking the CVC due to SC fluctuations into consideration, 
anomalous transport phenomena in HTSC's under $T^\ast$ 
(e.g., prominent enhancement of the Nernst coefficient) 
are well understood in a unified way
 \cite{Kontani-rev,Kontani-N}.

One of the merits of the FLEX approximation is that
the Mermin-Wagner theorem with respect to the magnetic 
instability is satisfied.
In fact, previous numerical studies based on the FLEX report 
that no SDW-order emerges in two dimensional systems at finite $T$.
In Appendix A, we offer an strong analytical evidence that the
FLEX approximation satisfies the Mermin-Wagner theorem.
Also, the $GV^I$-method does not predict any SDW-order
in two dimensional systems with single impurities, 
at least for model parameters studied in the present paper.
However, when the concentration of impurities is finite,
Swiss cheese holes will overlap when $\xi_{\rm AF}$ exceeds 
the mean distance between impurities ($l$).
Therefore, a SDW-order or a spin-glass order would happen
when $\xi_{\rm AF}>l$.
In fact, in Zn-doped YBCO,
the freezing of local moments due to Zn was 
observed by $\mu$-SR study at very low temperatures 
 \cite{mSR}.
Note that impurity induced AF order occurs 
in two-leg ladder Heisenberg models
 \cite{Fukuyama}.

\subsection{Comments on Related Theoretical Works}
\label{sec:comment}
Here, we discuss the impurity effect on the electronic states 
of the ``host system''.
We did not study this effect because it is higher order effect 
with respect to $n_{\rm imp}$.
In Zn-doped YBa$_2$Cu$_4$O$_8$,
Ihot et al. measured $^{63}$Cu $1/T_1T$ in the host system 
(i.e., away from the Zn sites) \cite{Itoh},
and found that the host AF fluctuations above $T^\ast$
are reduced by a few percent doping of Zn.
They suggest that the localization effect would be
the origin of this reduction.

Based on the FLEX approximation ($GV$-method in the present paper),
authors of ref. \cite{Kudo} studied the impurity effect on the electronic
state by neglecting the $\k$-dependence of the self-energy.
They reported that the AF fluctuation in the host system is depressed
by impurities, which seems to be consistent with ref. \cite{Itoh}.
As we have shown however, the $GV$-method fails to 
reproduce correct electronic states around the impurity:
In fact, both the Curie like $\Delta\chi^s$
as well as the huge $\Delta\rho$ are 
satisfactorily reproduced only by the $GV^I$-method.
We note that $\rho(T)$ in HTSC shows a nearly parallel shift
by the impurity doping, which means that the inelastic
scattering in the host system is independent of impurities.
This fact would suggest that the AF fluctuations in the host 
system is unchanged, in contrast to the NMR result \cite{Itoh}.
It would be an important issue to
understand these two experimental facts consistently.

Next, we comment that Miyake et al. have intensively studied the 
impurity effects near QCP
 \cite{Miyake02}.
They found that $\Delta\rho$ due to weak
impurities (Born scattering) is enlarged
when the charge fluctuations are developed.
Their analysis corresponds to $GV^0$-method
since the susceptibility is assumed to be independent of impurities.
In contrast, the present work based on $GV^I$-method
shows that the $\Delta\rho$ due to strong impurities
increases near the AF-QCP,
which originates from the enhancement of $\chi^{Is}$
around the impurity.
We note that $\chi^{\rm c}_{\rm FLEX} \equiv d n/d\mu$ 
is less than half of 
its non-interacting value in the present range of parameters.
In analysing experiments, we have to consider carefully 
which kind of criticality may occur in the compound under consideration.

Finally, we explain previous studies on disordered
Hubbard model based on the dynamical-mean-field-theory (DMFT) 
in the $d=\infty$ limit
 \cite{Kotliar,Vollhardt,Mutou},
and we make comparison with the present study.
In the DMFT, the local-moment formation is found
when strong hopping disorders (offdiagonal disorders) exist
 \cite{Kotliar}.
Also, it is found that the N{\' e}el temperature increases
due to weak onsite disorders (diagonal disorders); $I<U$
 \cite{Vollhardt}.
However, a local-moment formation outsside of the impurity
and a huge residual resistivity, which are realized
in HTSC's, cannot be explained by the DMFT.
For this purpose, a nonlocal modulation of the self-energy 
around the impurity have to be taken into account, 
which is possible in the $GV^I$-method.

\begin{acknowledgements}
We would like to thank
Y. Ando, Y. Matsuda, T. Shibauchi, T. Sekitani, M. Ido, M. Oda, 
N. Momono, H. Taniguchi, M. Sato, K. Yamada, D.S. Hirashima, 
Y. Tanaka, M. Ogata, Y. Yanase and S. Onari
for valuable comments and discussions.
\end{acknowledgements}

\appendix
\section{Mermin-Wagner theorem in 2D electron systems}
\label{sec:Ap}

The Mermin-Wagner (M-W) theorem states that
any magnetic instabilities are absent at finite $T$ in 2D systems.
Actually, the SCR theory satisfies the M-W theorem
 \cite{Moriya}.
As for the FLEX approximation, however, 
the M-W theorem had been recognized only by numerical studies.
In this appendix, we present a strong analytical evidence that the 
FLEX-type self-consistent spin fluctuation theory satisfies 
the M-W theorem.

Here we introduce the phenomenological expression for
the dynamical spin susceptibility as
\begin{eqnarray}
\chi_\q(\w) = \frac{\chi_{\bf Q}}{1+\xi_{\rm AF}^2(\q-\Q)^2-i\w/\w_{\rm sf}},
\end{eqnarray}
where $\xi_{\rm AF}$ is the AF correlation length.
$\Q$ is one of the nesting vectors which minimize $|\q-\Q|$.
Apparently, $\chi_\q(0)=\chi_{-\q}(0)$.
Both $\chi_{\bf Q}$ and $\w_{\rm sf}^{-1}$
are proportional to $\xi_{\rm AF}^2$ in the FLEX approximation.

When the system is very close to the AF phase
at finite temperatures where $\w_{\rm sf} \gg T$
is satisfied,
the self-energy within the 
scheme of the FLEX approximation is given by
\begin{eqnarray}
\Sigma_\k(i\w_n)
 &\approx& T\sum_\q G_{\k+\q}(i\w_n)\chi_\q(0)
 \nonumber \\
 &\approx& G_{\k+\Q}(i\w_n)A
 \label{eqn:S0}
 \\ 
A&=&T\frac{3U^2}{2}\sum_\q \chi_\q(0)
 \label{eqn:A}
\end{eqnarray}
where the static approximation is applied,
which will offer the upper limit of $T_N$.
$A$ diverges as $T\rightarrow T_N$
in proportion to $T\ln\xi_{\rm AF}$.
Especially, $A=\infty$ at $T=T_N(>0)$.
On the other hand,
$A$ is finite even at $T=T_N$ in 3D systems.
According to eq.(\ref{eqn:S0}),
\begin{eqnarray}
\Sigma_\k(i\w_n)
 &=& \frac{A}{i\w_n-\e_{\k+\Q}- \Sigma_{\k+\Q}(i\w_n)}
 \nonumber \\
 &=& \frac{A}{i\w_n-\e_{\k+\Q}- \frac{A}{i\w_n-\e_{\k}- \Sigma_{\k}(i\w_n)}}
 \label{eqn:S1}
\end{eqnarray}
Equation (\ref{eqn:S1}) can be solved analytically.
Considering $\Sigma_\k(\w)=0$ when $A=0$,
the self-energy and the Green function
for real frequencies are given by
\begin{widetext}
\begin{eqnarray}
\Sigma_\k(\w)
 &=& \frac12 (\w-\e_\k) - {\rm sgn}(\w-\e_\k)\frac12
 \sqrt{(\w-\e_\k)^2-4A\frac{\w-\e_\k}{\w-\e_{\k+\Q}}}
 \\
G_\k(\w)
 &=& \left(
\frac12 (\w-\e_\k) + {\rm sgn}(\w-\e_\k)\frac12
 \sqrt{(\w-\e_\k)^2-4A\frac{\w-\e_\k}{\w-\e_{\k+\Q}}}
\right)^{-1}
 \label{eqn:G1}
\end{eqnarray}
One can check that $\w\cdot{\rm Im}\Sigma_\k(\w) \le0$.
This is not a Fermi liquid because the renormalization factor 
$z$ is zero (owing to the static approximation).
Hereafter, we assume $\e_\k = 2t(\cos k_x+ \cos k_y)$
at half filling ($n=1$),
that is, both the perfect nesting and the particle-hole 
symmetry exist.
Apparently, $Q=(\pi,\pi)$,
$\e_{\k+\Q}=-\e_\k$, and $\mu=0$.
The irreducible susceptibility at $\q=\Q$ and $\w=0$
is given by
\begin{eqnarray}
\Pi_\Q(0) &=&
 -\sum_\k\int\frac{d\w}{2\pi} {\rm th}\frac{\w}{2T}
 {\rm Im} \{ G_{\k+\Q}^R(\w)G_{\k}^R(\w) \}
 \label{eqn:C1}
\end{eqnarray}
According to eq. (\ref{eqn:G1}),
\begin{eqnarray}
G_{\k+\Q}(\w)G_{\k}(\w)
&=& \left(\frac14 (\w^2-\e_\k^2)
 + \frac12 {\rm sgn}(\w^2-\e_\k^2)\sqrt{g_\k(\w)}
 +\frac14 {\rm sgn}(\w^2-\e_\k^2)
 |\w^2-\e_\k^2-4A| \right)^{-1}
 \\
g_\k(\w)
&=& (\w^2-\e_\k^2)(\w^2-\e_\k^2-4A)
\end{eqnarray}
$g_\k(\w)$ is negative when $|\e_\k| < |\w| < \sqrt{\e_\k^2+4A}$.
Apparently, the integrand in eq.(\ref{eqn:C1})
is finite only when $g_\k(\w)<0$.
\begin{eqnarray}
-{\rm Im} \{ G_{\k+\Q}(\w)G_{\k}(\w) \}
&=& -{\rm Im} \{ ( \ A+\frac{i}{2} \sqrt{-g_\k(\w)} \ )^{-1} \}
< \frac{1}{2A} \ \ \ \ \mbox{for $g_\k(\w)<0$}
 \\
&=& 0 \ \ \ \ \ \ \ \ \ \ \ \ \ \ \ \ \ \ \ \ \ \ \ \ \ \ \ \ 
 \ \ \ \ \ \ \ \ \ \ \ \ \ \ \ \ \ \ \ \
\mbox{for $g_\k(\w)>0$} 
\end{eqnarray}
\end{widetext}
According to eq.(\ref{eqn:C1}),
in the case of $\sqrt{A} \gg W_{\rm band}$,
\begin{eqnarray}
\Pi_\Q(0) \sim O(A^{-1/2})
 \label{eqn:chi}
\end{eqnarray}
Note that 
Im$G_\k(\w)$ is non-zero when $g_\k(\w)<0$, and
one can check that
$\int_{-\infty}^\infty d\w {\rm Im} G_\k^R(\w)=-\pi$
for any $A$ by MATHEMATICA.
Considering that $\chi_\Q(0)=\Pi_\Q(0)/(1-U\Pi_\Q(0))$
in the FLEX approximation,
eq. (\ref{eqn:chi})
means that $\chi_\Q(0)$ approaches to zero
as $T \rightarrow T_N$ (because $A\rightarrow\infty$ 
in the case of $T_N>0$) in 2D systems.
Equation (\ref{eqn:chi}) suggests that
$\xi_{\rm AF} \propto e^{1/T}$  
when the ground state is a ordered state.

As a result, $T_N$ cannot take a finite value in the
two dimensional model with perfect nesting at half filling,
which is the most likely to cause the magnetic instability.
Therefore, the present analysis gives a compelling evidence
the the FLEX approximation satisfies the M-W theorem
for general 2D systems.

Here we rewrite $A$ as $A'+A''$, where $\displaystyle 
A'=T \frac{3U^2}{2}\sum_\q^{|\q-{\bf Q}|<q_{\rm c}} \chi_\q(0)$,
and $q_{\rm c}$ is a cutoff momentum.
Then, $A''$ is a smooth function around the AF-QCP.
When $T>\xi_{\rm AF}^{-2}+q_{\rm c}^2$,
$A'$ will be the number of free bosons (magnons) with energy 
$\e_\q= q^2$ within the radius of $q_{\rm c}$ 
at the chemical potential $\mu=-\xi_{\rm AF}^{-2}$.
Thus, a magnetic instability ($\xi_{\rm AF}\rightarrow\infty$)
corresponds to the Bose-Einstein condensation of magnons ($\mu=0$).
Therefore, a meaningful correspondence between the M-W theorem
and the absence (presence) of the Bose-Einstein condensation 
in 2D (3D) systems is recognized.



\end{document}